\documentclass[12pt]{article}

\usepackage[a4paper,text={16.8cm,22.4cm}]{geometry}
\usepackage{amsmath,amsfonts,slashed,amssymb,tikz,bm,psfrag,graphicx,color,dsfont}
\usepackage{multicol}

\RequirePackage[sort&compress,square,comma,numbers]{natbib}
\allowdisplaybreaks 
\begin{document}

\begin{titlepage}

\begin{flushright}
\normalsize
MITP/15-002\\ 
January 26, 2015\\
arXiv:1501.06569
\end{flushright}

\vspace{0.1cm}
\begin{center}
\Large\bf\boldmath
Exclusive Radiative Decays of $W$ and $Z$ Bosons\\ 
in QCD Factorization
\end{center}

\vspace{0.5cm}
\begin{center}
Yuval Grossman$^a$, Matthias K\"onig$^b$ and Matthias Neubert$^{a,b}$\\
\vspace{0.7cm} 
${}^a$Department of Physics, LEPP, Cornell University, Ithaca, NY 14853, U.S.A.\\[3mm]
{\sl ${}^b$PRISMA Cluster of Excellence \& Mainz Institute for Theoretical Physics\\
Johannes Gutenberg University, 55099 Mainz, Germany}
\end{center}

\vspace{0.8cm}
\begin{abstract}
We present a detailed theoretical analysis of very rare, exclusive hadronic decays of the electroweak gauge bosons $V=W,Z$ from first principles of QCD. Our main focus is on the radiative decays $V\to M\gamma$, in which $M$ is a pseudoscalar or vector meson. At leading order in an expansion in powers of $\Lambda_{\rm QCD}/m_V$ the decay amplitudes can be factorized into convolutions of calculable hard-scattering coefficients with the leading-twist light-cone distribution amplitude of the meson $M$. Power corrections to the decay rates arise first at order $(\Lambda_{\rm QCD}/m_V)^2$. They can be estimated in terms of higher-twist distribution amplitudes and are predicted to be tiny. We include one-loop ${\cal O}(\alpha_s)$ radiative corrections to the hard-scattering coefficients and perform the resummation of large logarithms $\big(\alpha_s\ln(m_V^2/\mu_0^2)\big)^n$ (with $\mu_0\sim 1$\,GeV a typical hadronic scale) to all orders in perturbation theory. Evolution effects have an important impact both numerically and conceptually, since they reduce the sensitivity to poorly determined hadronic parameters. We present detailed numerical predictions and error estimates, which can serve as benchmarks for future precision measurements. We also present an exploratory study of the weak radiative decays $Z\to MW$. Some of the decay modes studied here have branching ratios large enough to be accessible in the high-luminosity run of the LHC. Many of them can be measured with high accuracy at a future lepton collider. This will provide stringent tests of the QCD factorization formalism and enable novel searches for new physics. 
\end{abstract}
\vfil

\end{titlepage}

\section{Introduction}

One of the main challenges to particle physics is to obtain a rigorous control of strong-interaction phenomena in a regime where QCD is strongly coupled. Over the years, lattice QCD has made much progress in computing the static properties of hadrons from first principles. The concept of quark-hadron duality has enabled us to make systematic predictions for inclusive decay processes with a large energy release, such as $e^+ e^-\to\mbox{hadrons}$ at large $\sqrt{s}$, or inclusive weak decays like $B\to Xl\nu$. In these cases, non-perturbative aspects of the strong interactions can be accounted for using a local operator-product expansion. A conceptually more difficult problem is to control strong-interaction effects in exclusive hadronic processes at large energy. For deep-inelastic scattering a factorization theorem can be derived, in which all non-perturbative physics associated with the initial-state nucleon can be described in terms of parton distribution functions (PDFs), up to power corrections suppressed by $\Lambda_{\rm QCD}/\sqrt{s}$. The same framework is routinely used to calculate cross sections at hadron colliders such as the LHC in terms of convolutions of calculable partonic cross sections with PDFs, even though the underlying factorization formula can only be proved for the simplest such processes. 

The QCD factorization approach developed by Brodsky and Lepage \cite{Lepage:1979zb,Lepage:1980fj}, Efremov and Radyushkin \cite{Efremov:1978rn,Efremov:1979qk} and others \cite{Chernyak:1983ej} provides a theoretical basis for controlling strong-interaction effects in exclusive processes with individual, highly energetic hadrons in the final state. Bound-state effects are accounted for in terms of light-cone distribution amplitudes (LCDAs) of these hadrons, which are defined in terms of the matrix elements of non-local quark and gluon operators with light-like separation. This approach provides an expansion of amplitudes in powers of $\Lambda_{\rm QCD}/Q$, where $Q$ is the large energy released to the hadronic final state. While the leading term can be calculated in a model-independent way, it is generally not guaranteed that power corrections can be meaningfully computed, as they may involve ill-defined overlap integrals. About 15 years ago the QCD factorization formalism was generalized to deal with a particularly complicated class of processes: non-leptonic, exclusive weak decays of $B$ mesons \cite{Beneke:1999br,Beneke:2000ry,Beneke:2001ev,Beneke:2003zv}. The additional complication consists in the presence of soft form-factor contributions in addition to the hard-scattering contributions described in the Brodsky-Lepage framework. The decay amplitudes are expanded in powers of $\Lambda_{\rm QCD}/E$, where $E\sim m_b$ is the energy of the final-state hadrons in the $B$-meson rest frame. While the predictions obtained at leading order are theoretically clean and in reasonable agreement with experiment, already the first-order power corrections involve ill-defined overlap integrals. This introduces poorly known model parameters and makes phenomenological predictions less precise. Since the advance of soft-collinear effective theory (SCET) \cite{Bauer:2000yr,Bauer:2001yt,Bauer:2002nz,Beneke:2002ph}, the QCD factorization approach can be rephrased in the language of effective field theory, which helps making its workings more transparent.

All existing applications of the QCD factorization approach have suffered from the fact that the characteristic energy scales are not sufficiently large for power corrections to be negligible (see e.g.\ \cite{Agaev:2010aq} for a recent discussion). It is then notoriously difficult to disentangle power-suppressed effects from the uncertainties related to the shapes of the hadron LCDAs. Unfortunately, no comprehensive experimental program to determine the leading-twist LCDAs of the ground-state mesons and baryons -- analogous to the large-scale effort to determine the PDFs of the proton with high accuracy -- is conceivable. In this paper we propose using exclusive decays of the heavy electroweak gauge bosons $W$ and $Z$ into final states containing a single meson as a laboratory to test and study the QCD factorization approach in a context where power corrections are definitely under control. The enormous rates of $W$ and $Z$ bosons that will become available at future colliders will present us with a new playground for precision electroweak and QCD physics, which will make such studies feasible. With 3000\,fb$^{-1}$ collected during the high-luminosity run at the LHC, one will have produced more than $10^{11}$ $Z$ bosons and $5\cdot 10^{11}$ $W$ bosons in both ATLAS and CMS. A clean, tagged sample of $W$ bosons is expected to come from top-quark decays \cite{Mangano:2014xta}. At a future lepton collider such as TLEP, samples of up to $10^{12}$ $Z$ bosons per year can be expected in a dedicated run at the $Z$ pole \cite{Blondel:2013rn}. Our main focus is on the simplest processes: the hadronic radiative decays $Z\to M\gamma$ and $W\to M\gamma$, where $M$ is a pseudoscalar or vector meson. They offer a perfect way to probe some properties of the leading-order LCDAs of various mesons. The price one needs to pay is that the higher the energy release in the process, the smaller the probabilities for any particular exclusive final state are. The branching fractions we obtain range from few times $10^{-8}$ to few times $10^{-11}$ or even smaller.  The big challenge of such a program will be to measure such decays experimentally with some precision. While we do not perform a detailed feasibility study in this work, we speculate that some of these rare modes will be accessible in the high-luminosity run of the LHC, at a level that will be useful to probe our theoretical predictions. At future lepton colliders operating on the $Z$ pole, it would be possible to measure several of these decays at or below the 1\% level. This may present us with a unique opportunity to extract information about LCDAs in a theoretically clean environment. 

Our interest in this subject was raised by recent investigations of the exclusive decays $h\to V\gamma$ \cite{Isidori:2013cla,Bodwin:2013gca,Kagan:2014ila,Bodwin:2014bpa} and $h\to Z V$ \cite{Gao:2014xlv,Bhattacharya:2014rra} of the Higgs boson to final states containing a single vector meson. It was proposed to use these decays as a way to probe for possible non-standard Yukawa couplings of the Higgs boson. Such measurements are extremely challenging at the LHC and other future colliders. Observing exclusive hadronic decays of $W$ and $Z$ bosons would provide a proof of principle that this kind of searches can be performed. An encouraging first search for the decays $Z\to J/\psi\,\gamma$ and $Z\to \Upsilon(nS)\,\gamma$ has just been reported by ATLAS \cite{Aad:2015sda}.

From a theoretical perspective, the very rare, exclusive radiative decays of $W$ and $Z$  bosons have received relatively little attention in the literature, and very few accurate predictions for such branching fractions have been obtained. In a pioneering study \cite{Arnellos:1981gy}, Arnelos, Marciano and Parsa presented a first detailed analysis of the decays $W\to P\gamma$ and $Z\to P\gamma$ for both light and heavy pseudoscalar mesons in the final state. Strong-interaction effects were parametrized in terms of vector and axial-vector form factors, which were estimated using ideas from perturbative QCD on the asymptotic behavior of form factors at large momentum transfer. Several years later, Manohar studied the decays $Z\to\pi W$ and $Z\to\pi\gamma$ using a local operator-product expansion \cite{Manohar:1990hu}, which expresses the decay amplitudes as power series in parameters $\omega_0=\frac{2(m_Z^2-m_W^2)}{m_Z^2+m_W^2}\approx 0.26$ and $\omega_0=2$, respectively. If only the leading term is kept, the amplitudes can be related to the pion matrix element of the axial-vector current, which is proportional to the pion decay constant $f_\pi$. For radiative decays such as $Z\to\pi\gamma$ this truncation cannot be justified theoretically, and the infinite tower of local operators would need to be resummed. In a very recent work, the method developed by Manohar was used to derive an estimate for the $W\to\pi\gamma$ branching fraction \cite{Mangano:2014xta}. While the analyses performed in these papers can provide some order-of-magnitude results, they do not allow to obtain accurate predictions with reliable error estimates. In a classic paper \cite{Guberina:1980dc}, Guberina {\em et al.\/} analyzed the radiative decays of the $Z$ boson into heavy quarkonia in the non-relativistic limit. The first relativistic corrections to the $Z\to J/\psi\,\gamma$ and $Z\to\Upsilon(1S)\,\gamma$ decay rates were added only recently in \cite{Huang:2014cxa}, where in addition the authors considered for the first time the decay $Z\to\phi\gamma$, using an approach closely related to ours. As we will discuss, renormalization effects have a profound impact on the decay amplitudes. When evolved up to the relevant scales of order the $Z$-boson mass, the LCDAs of heavy quarkonia can no longer be accurately described by the leading term in a non-relativistic expansion.

In the present work, we present a comprehensive analysis of a large class of radiative decays of $W$ and $Z$ bosons using the QCD factorization approach, including for the first time a consistent treatment of ${\cal O}(\alpha_s)$ corrections and performing the resummation of large logarithms of order $\big(\alpha_s\ln(m_Z^2/\mu_0^2)\big)^n$, with $\mu_0\approx 1$\,GeV, to all orders in perturbation theory.\footnote{Radiative corrections to the $Z\to J/\psi\,\gamma$ and $Z\to\Upsilon(1S)\,\gamma$ decay amplitudes were included in \cite{Huang:2014cxa} in the non-relativistic limit, but they were not included so far in any analysis of decays into light final-state mesons.} 
Our approach provides a systematic expansion of the decay amplitudes in powers of the small parameters $\alpha_s(m_Z)\sim 0.1$ and $\Lambda_{\rm QCD}/m_Z\sim 0.01$. We study the structure of the leading power corrections to the $Z\to M\gamma$ and $W\to M\gamma$ decay rates and show that they are of second order and hence negligibly small, of order $10^{-4}$ relative to the leading terms. For processes involving heavy quarks, power corrections of order $(m_Q/m_Z)^2$ exist, which are still very small (less than 1\%) even for final-state mesons containing $b$-quarks. Finally, using the most recent experimental data we perform a reanalysis of meson decay constants, which provide crucial input to our phenomenological analysis.

Our paper is organized as follows: In Section~\ref{sec:theory} we derive a factorization theorem for the $Z\to M\gamma$ and $W\to M\gamma$ decay amplitudes, in which they are expressed as convolutions of calculable hard-scattering kernels with meson LCDAs. We explain how the kernel functions can be calculated by performing projections of on-shell partonic amplitudes. We then summarize the existing theoretical information on the shapes of the LCDAs for both light and heavy mesons and study their behavior under scale evolution. In Section~\ref{sec:raddecays} we apply this approach to derive explicit predictions for the $Z\to M\gamma$ and $W\to M\gamma$ decay amplitudes at leading power in $\Lambda_{\rm QCD}/m_Z$. The relevant convolution integrals of hard-scattering kernels with LCDAs are calculated in analytic form using an expansion in Gegenbauer polynomials. We demonstrate that renormalization-group (RG) evolution from a low hadronic scale up to the electroweak scale of relevance to these processes has the nice effect of significantly reducing the sensitivity to poorly determined hadronic parameters. By studying radiative decays into transversely polarized vector mesons, we present some detailed estimates of power-suppressed effects in the QCD factorization approach. In some old papers, it was suggested that the radiative decay amplitudes into pseudoscalar mesons can be hugely enhanced due to effects of the axial anomaly \cite{Jacob:1989pw,Keum:1993eb}. We explain why such an enhancement does not exist. We then present our numerical predictions for $W,Z\to M\gamma$ branching fractions, including detailed error estimates. The results span more than three orders of magnitude, and we explain the striking differences seen between the various decay channels in terms of electroweak couplings, differences in decay constants, and enhancement factors occurring for heavy-light mesons. In Section~\ref{sec:weakrad} we present a first exploratory study of the weak radiative decays $Z\to MW$, in which a heavy $W$ boson is part of the final state. In this case significantly less energy is released to the final-state meson $M$, and as a result the QCD factorization approach can be tested at energies of order 10\,GeV, about a factor~2 higher than those available in exclusive $B$-meson decays. We round off our study in Section~\ref{sec:experiments} with some experimental considerations. Our main results are summarized in Section~\ref{sec:concl}. Technical details of our calculations and the extraction of meson decay constants are relegated to three appendices.

\section{Theoretical framework}
\label{sec:theory}

\begin{figure}
\begin{center}
\includegraphics[width=0.3\textwidth]{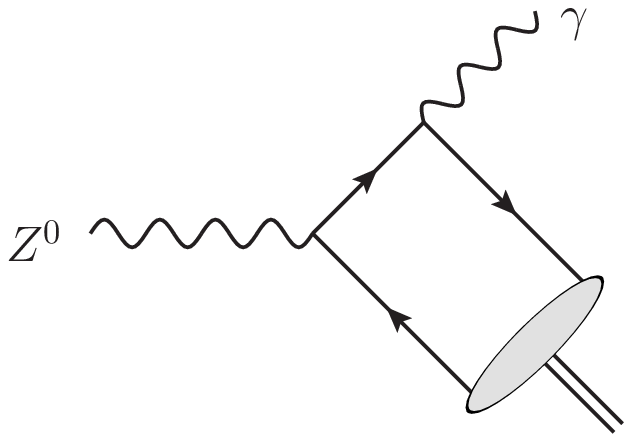}
\includegraphics[width=0.3\textwidth]{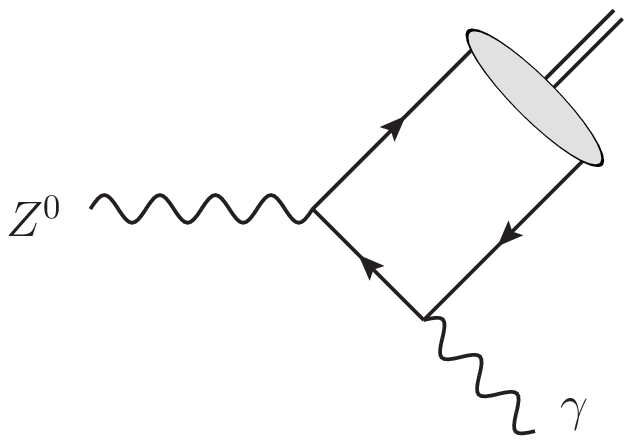}
\parbox{15.5cm}
{\caption{\label{fig:LOgraphs}
Leading-order Feynman diagrams for the radiative decays $Z^0\to M^0\gamma$. The meson bound state is represented by the gray blob.}}
\end{center}
\end{figure}

Our main focus in this work is on the rare, exclusive radiative decays $Z\to M\gamma$ and $W\to M\gamma$, where $M$ denotes a pseudoscalar or vector meson. We assign momentum $k$ to the final-state meson and $q$ to the photon. The leading-order Feynman diagrams for the case of $Z\to M\gamma$ are shown in Figure~\ref{fig:LOgraphs}. The decay plane is spanned by the vectors $k$ and $q$. We will refer to vectors in this plane as being longitudinal, and to vectors orthogonal to it as being transverse. We only consider cases where the mass of the final-state meson satisfies $m_M\ll m_Z$. Up to corrections suppressed as $(m_M/m_Z)^2$, this mass can then be set to zero. In this limit, we have $k^\mu=E n^\mu$ and $q^\mu=E\bar n^\mu$, where $E=m_Z/2$ is the energy of the final-state particles in the $Z$-boson rest frame, and $n$ and $\bar n$ are two light-like vectors satisfying $n\cdot\bar n=2$.

\subsection{Derivation of the factorization formula}

For the purposes of this discussion we work in the rest frame of the decaying heavy boson. The decay amplitudes can be calculated from first principles using the QCD factorization approach \cite{Lepage:1979zb,Lepage:1980fj,Efremov:1978rn,Efremov:1979qk,Chernyak:1983ej}, because the energy $E$ released to the final-state meson is much larger than the scale of long-distance hadronic physics. At leading power in an expansion in $\Lambda_{\rm QCD}/m_Z$, they can be written as convolutions of calculable hard-scattering coefficients with LCDAs of the meson $M$. A simple way to derive the corresponding factorization theorem employs the formalism of SCET \cite{Bauer:2000yr,Bauer:2001yt,Bauer:2002nz,Beneke:2002ph}. It provides a systematic expansion of decay amplitudes in powers of a small expansion parameter $\lambda=\Lambda_{\rm QCD}/E$. The light final-state meson moving along the direction $n^\mu$ can be described in terms of collinear quark, anti-quark and gluon fields. These particles carry collinear momenta $p_c$ that are approximately aligned with the direction $n$. Their components scale like $(n\cdot p_c, \bar n\cdot p_c, p_c^\perp)\sim E(\lambda^2,1,\lambda)$. Note that $p_c^2\sim\Lambda_{\rm QCD}^2$, as appropriate for an exclusive hadronic state. The collinear quark and gluon fields are introduced as gauge-invariant objects dressed with Wilson lines. Explicitly, one defines \cite{Bauer:2001ct,Hill:2002vw}
\begin{equation}\label{SCETfields}
   {\cal X}_c = \frac{\rlap{\hspace{0.02cm}/}{n}\rlap{\hspace{0.02cm}/}{\bar n}}{4}\,
    W_c^\dagger\,q \,, \qquad
   {\cal A}_{c\perp}^\mu = W_c^\dagger\,(iD_{c\perp}^\mu W_c) \,, 
\end{equation}
where $iD_c^\mu=i\partial^\mu+ig A_c^\mu$ denotes the covariant collinear derivative, and
\begin{equation}
   W_c(x) = {\bf P} \exp\left( ig\int_{-\infty}^0\!dt\,\bar n\cdot A_c(x+t\bar n) \right)
\end{equation}
is a collinear Wilson line extending from $x$ to infinity along the direction $\bar n$. Both fields are of ${\cal O}(\lambda)$ in SCET power counting. Adding more component fields to an operator always leads to further power suppression. At leading order in $\lambda$, the operators with a non-zero matrix element between the vacuum and a single meson state are thus of the form ${\bar{\cal X}}_c(t\bar n)\,\dots\,{\cal X}_c(0)$ and ${\cal A}_{c\perp}^\mu(t\bar n)\,\dots\,{\cal A}_{c\perp\mu}(0)$, where without loss of generality we set $x=0$ for one of the fields. Since the effective collinear fields are gauge invariant by themselves, composite operators built out of these fields can be non-local along the light-like direction $\bar n$. The two-gluon operator would only be relevant for decays into mesons containing a flavor-singlet component on their wave functions, such as the pseudoscalar mesons $\eta$ and $\eta'$ \cite{Beneke:2002jn}. Such decays will be discussed in a forthcoming publication \cite{inprep}. It follows that at leading power in the expansion in $\lambda$, the $Z\to M\gamma$ and $W\to M\gamma$ decay amplitudes into non-singlet final states can be written in the factorized form
\begin{equation}
\begin{aligned}
   {\cal A} 
   &= \sum_i \int dt\,C_i(t,\mu)\,
    \langle M(k)|\,{\bar{\cal X}}_c(t\bar n)\,\frac{\rlap{\hspace{0.02cm}/}{\bar n}}{2}\,
    \Gamma_i\,{\cal X}_c(0) |0\rangle + \mbox{power corrections} \\
   &= \sum_i \int dt\,C_i(t,\mu)\,
    \langle M(k)|\,\bar q(t\bar n)\,\frac{\rlap{\hspace{0.02cm}/}{\bar n}}{2}\,
    \Gamma_i\,[t\bar n,0]\,q(0) |0\rangle + \mbox{power corrections,}
\end{aligned}
\end{equation}
where $\mu$ is the factorization scale, and $\Gamma_i\in\{1,\gamma_5,\gamma^\mu_\perp\}$. The four matrices $(\rlap{\hspace{0.02cm}/}{\bar n}/2)\,\Gamma_i$ provide a basis of Dirac matrices sandwiched between two collinear quark spinors. The Wilson coefficients $C_i(t)$ are process dependent and can be calculated perturbatively. In the last step we have used the definition (\ref{SCETfields}) and combined the two Wilson lines $W_c(t\bar n)\,W_c^\dagger(0)\equiv[t\bar n,0]$ into a straight Wilson line extending from $0$ to $t\bar n$. The meson matrix elements of the bi-local operators in the second line define the leading-order LCDAs of pseudoscalar and vector mesons. Specifically, one has 
\begin{equation}
\begin{aligned}
   \langle M(k)|\,\bar q(t\bar n)\,\frac{\rlap{\hspace{0.02cm}/}{\bar n}}{2}\,(\gamma_5)\,
    [t\bar n,0]\,q(0) |0\rangle
   &= - if_M E \int_0^1\!dx\,e^{ixt\bar n\cdot k}\,\phi_M(x,\mu) \,; 
    \quad M=P,V_\parallel \,, \\
   \langle V_\perp(k)|\,\bar q(t\bar n)\,
    \frac{\rlap{\hspace{0.02cm}/}{\bar n}}{2}\,\gamma_\perp^\mu\,[t\bar n,0]\,q(0) |0\rangle
   &= - if_V^\perp(\mu) E\,\varepsilon_V^{\perp *\mu} \int_0^1\!dx\,e^{ixt\bar n\cdot k}\,
    \phi_V^\perp(x,\mu) \,,
\end{aligned}
\end{equation}
where $E=\bar n\cdot k/2$ denotes the energy of the meson in the rest frame of the decaying boson, $f_P$ and $f_V$ are the decay constants of pseudoscalar and vector mesons defined in terms of their matrix elements of local (axial-)vector currents, and $f_V^\perp(\mu)$ is a scale-dependent vector-meson decay constant defined in terms of a matrix element of the QCD tensor current. The leading-order LCDAs can be interpreted as the amplitudes for finding a quark with longitudinal momentum fraction $x$ insinde the meson. The factor of $\gamma_5$ in the first equation is present for a pseudoscalar meson ($M=P$) but absent for a longitudinally polarized vector meson ($M=V_\parallel$). The projection onto a transversely polarized vector meson does not arise at leading power in the radiative decays of $W$ and $Z$ bosons. For a given meson, exactly one of the possible Dirac structures contributes, and we denote the corresponding Wilson coefficient by $C_M(t,\mu)$. Defining the Fourier-transformed Wilson coefficient, called the hard function, via
\begin{equation}
   H_M(x,\mu)\equiv\int dt\,C_M(t,\mu)\,e^{ixt\bar n\cdot k} \,,
\end{equation}
we obtain the factorization formula
\begin{equation}\label{factorization}
   {\cal A} = -i f_M E \int_0^1\!dx\,H_M(x,\mu)\,\phi_M(x,\mu) + \mbox{power corrections} \,.
\end{equation}
Insertions of additional collinear fields or derivatives yield power-suppressed contributions. In particular, the insertion of an additional collinear gluon field gives rise to three-particle LCDAs. In order to fully establish the factorization theorem (\ref{factorization}) one must show that the convolution integral over the momentum fraction $x$ converges at the endpoints. This question has been addressed in the context of the more complicated processes $B\to\gamma l\nu$ \cite{Bosch:2003fc} and $B\to K^*\gamma$ in \cite{Becher:2005fg}. The behavior near the endpoints can be described by means of soft-collinear fields \cite{Becher:2003qh,Becher:2003kh} with momenta scaling as $(n\cdot p_{sc},\bar n\cdot p_{sc},p_{sc}^\perp)\sim E(\lambda^2,\lambda,\lambda^{3/2})$. The contributions of such modes are always power suppressed. In the present case, we find that endpoint singularities are absent at leading and subleading power in the large-energy expansion. 

LCDAs play the same role for hard exclusive processes which PDFs play for inclusive ones. While they encode genuinely non-perturbative hadronic physics, they can be rigorously defined in terms of non-local operator matrix elements in QCD \cite{Lepage:1979zb,Lepage:1980fj,Efremov:1978rn,Efremov:1979qk,Chernyak:1983ej}. These matrix elements can be systematically expanded in terms of structures of different twist. When applied to high-energetic exclusive processes such as the ones considered here, the twist expansion translates into an expansion in powers of $\Lambda_{\rm QCD}/E$. There is an extensive amount of literature devoted to the study of distribution amplitudes. For light pseudoscalar mesons, the two- and three-particle LCDAs up to twist-3 order were studied, e.g., in \cite{Braun:1989iv}, while the corresponding LCDAs for vector mesons were analyzed, e.g., in \cite{Ali:1993vd,Ball:1996tb,Ball:1998sk}. We stress that, at the scale of the large energies released in decays of $W$ and $Z$ bosons, even charm and bottom quarks can be treated as light quarks, and hence heavy mesons containing these quarks can be described by LCDAs. This will be discussed further below.

In order to apply these results in practical calculations, it is convenient to define momentum-space projection operators, which can be applied directly to the decay amplitudes computed with on-shell external parton states \cite{Beneke:2000wa,Beneke:2001ev}. For all two-particle projections onto LCDAs of leading and subleading twist, it is sufficient to assign momenta $k_1=xk+k_\perp+\dots$ and $k_2=(1-x)k-k_\perp+\dots$ to the quark and the anti-quark in the meson $M$, where $k$ is treated as a light-like vector ($k^2=0$). Meson mass effects of order $m_M^2$ enter only at twist-4 level. They have a tiny numerical impact for the decays considered here, and we will consistently set $m_M^2\to 0$ unless noted otherwise. The variables $x$ and $(1-x)$ denote the longitudinal momentum fractions carried by the quark and the anti-quark in the two-body Fock state of the meson. Each Feynman diagram gives an expression of the form
\begin{equation}
\begin{aligned}
   \bar u(k_1)\,A(q,k_1,k_2)\,v(k_2) 
   &= \mbox{Tr}\left[ v(k_2)\,\bar u(k_1)\,A(q,k_1,k_2) \right] \\
   &\to \int_0^1\!dx\,\mbox{Tr}\left[ M_M(k,x,\mu)\,A(q,k_1,k_2) \right]_{k_\perp\to 0} ,
\end{aligned}
\end{equation}
where in the last step we have introduced the light-cone projection operator $M_M(k,x,\mu)$ for the meson $M$, which at higher order contains derivatives with respect to the parton transverse momentum $k_\perp$. It is understood that $k_\perp$ is set to zero after these derivatives have been performed. 

Up to twist-3 order, the light-cone projector for a pseudoscalar meson can be written in the form \cite{Beneke:2000wa,Beneke:2001ev} 
\begin{equation}\label{LCDAP}
\begin{aligned}
   M_P(k,x,\mu) &= \frac{if_P}{4}\,\Bigg\{ \rlap{\hspace{0.1mm}/}{k}\gamma_5\,\phi_P(x,\mu) 
    - \mu_P(\mu)\,\gamma_5 \bigg[ \phi_p(x,\mu) 
    - i\sigma_{\mu\nu}\,\frac{k^\mu\,\bar n^\nu}{k\cdot\bar n}\,
    \frac{\phi_\sigma'(x,\mu)}{6} \\[-1mm]
   &\hspace{15mm}\mbox{}+ i\sigma_{\mu\nu} k^\mu\,\frac{\phi_\sigma(x,\mu)}{6}\,
    \frac{\partial}{\partial k_{\perp\nu}} \bigg] + \mbox{3-particle LCDAs} \Bigg\} \,.
\end{aligned}
\end{equation}
Here $\phi_P$ is the leading-twist LCDA of the meson, while $\phi_p$ and $\phi_\sigma$ denote the two-particle LCDAs appearing at twist-3 order. These are scale-dependent functions, which we define in the $\overline{\rm MS}$ renormalization scheme. The decay constant $f_P$ of the meson $P$ is defined in terms of its matrix element of a local axial-vector current
\begin{equation}\label{fPdef}
   \langle P(k)|\,\bar q_1\gamma^\mu\gamma_5 q_2\,|0\rangle = -if_P k^\mu \,.
\end{equation}
The scale-dependent parameter $\mu_P(\mu)=m_P^2/[m_{q_1}(\mu)+m_{q_2}(\mu)]$ governs the normalization of the twist-3 LCDAs.\footnote{Note that $\mu_\pi=m_\pi^2/(m_u+m_d)$ holds for charged and neutral pions, see e.g.\ \cite{Beneke:2002jn}.} 
The vector $\bar n$ in the above expression denotes a longitudinal light-like vector not aligned with $k$. A convenient choice is to take the photon momentum, $\bar n=q$. At twist-3 order the projector also contains three-particle LCDAs containing a quark, an anti-quark and a gluon. We will see that the contributions of twist-3 LCDAs are strongly suppressed compared with those of the leading-twist amplitudes. In order to estimate their effects, we will for simplicity neglect the three-particle LCDAs. This is referred to as the Wandzura-Wilczek approximation (WWA) \cite{Wandzura:1977qf}. When this is done, the QCD equations of motion fix the form of the twist-3 LCDAs completely, and one obtains \cite{Braun:1989iv}
\begin{equation}
   \phi_p(x,\mu)\big|_{\rm WWA} = 1 \,, \qquad
   \phi_\sigma(x,\mu)\big|_{\rm WWA} = 6x(1-x) \,.
\end{equation}
The light-cone projection operators for vector mesons are more complicated. They are given in Appendix~\ref{app:LCDAs}. For our purposes it suffices to quote the projector for a longitudinally polarized vector meson at leading power. It is
\begin{equation}\label{LCDAV}
   M_{V_\parallel}(k,x,\mu) = - \frac{if_V m_V}{4}\,
    \frac{\varepsilon_V^{\parallel *}\!\cdot\bar n}{k\cdot\bar n}\,
    \rlap{\hspace{0.3mm}/}{k}\,\phi_V(x,\mu) + \dots
    = - \frac{if_V}{4}\,\rlap{\hspace{0.3mm}/}{k}\,\phi_V(x,\mu) + \dots \,.
\end{equation}
The function $\phi_V(x,\mu)$ is sometimes called $\phi_V^\parallel(x,\mu)$ in the literature. We have used that the longitudinal polarization vector is given by $\varepsilon_V^{\parallel\,\mu}=\frac{1}{m_V}\big(k^\mu-m_V^2\,\frac{\bar n^\mu}{k\cdot\bar n}\big)$. The vector-meson decay constant $f_V$ is defined in terms of the local matrix element
\begin{equation}\label{vectorme}
   \langle V(k,\varepsilon_V)|\,\bar q_1\gamma^\mu q_2\,|0\rangle 
   = -if_V m_V \varepsilon_V^{*\mu} \,.
\end{equation}

Before proceeding, let us comment on the structure of power corrections to the factorization formula (\ref{factorization}). Inspecting the explicit form of the projection operator for a pseudoscalar meson in (\ref{LCDAP}), and the corresponding projectors for vector mesons given in (\ref{LCDAVlong}) and (\ref{LCDAVperp}), we observe that consecutive terms in the twist expansion contain even and odd numbers of Dirac matrices in alternating order. Since the gauge interactions in the Standard Model preserve chirality, it follows that for a given helicity amplitude either all terms with an even number of Dirac matrices contribute or all terms containing an odd number, but not both. Consequently, the SCET expansion for the $Z\to M\gamma$ decay amplitudes with fixed polarizations of all particles is an expansion in powers of $(\Lambda_{\rm QCD}/m_Z)^2$. The power counting changes when quark-mass effects are taken into account. They give rise to chirality-changing vertices, which give corrections suppressed by $m_Q/m_Z$ to both the amplitudes and the meson projectors. This leads to power corrections of order $m_Q\Lambda_{\rm QCD}/m_Z^2$ and $(m_Q/m_Z)^2$. For heavy quarks with $m_Q\gg\Lambda_{\rm QCD}$, the latter corrections are the dominant ones. However, as long as the relevant quark masses $m_Q$ are much smaller than the hard scale $m_Z$ of the process, these corrections are still small. The present case is different from the situation encountered in exclusive $B$-meson decays \cite{Beneke:1999br,Beneke:2000ry,Beneke:2001ev,Beneke:2003zv}, where the presence of a heavy quark mass, which is of the same order as the energy released in the decay, allows for ${\cal O}(1)$ chirality-changing interactions. In this case the decay amplitudes receive first-order $\Lambda_{\rm QCD}/m_b$ corrections.

\subsection{Systematics of the Gegenbauer expansion}

The leading-twist LCDAs obey an expansion in Gegenbauer polynomials of the form \cite{Lepage:1979zb,Chernyak:1983ej}
\begin{equation}\label{Gegenbauer}
   \phi_M(x,\mu) = 6x(1-x) \left[ 1 + \sum_{n=1}^\infty a_n^M(\mu)\,C_n^{(3/2)}(2x-1) \right] ,
\end{equation}
which can be inverted to give
\begin{equation}
   a_n^M(\mu) = \frac{2(2n+3)}{3(n+1)(n+2)}\,\int_0^1\!dx\,C_n^{(3/2)}(2x-1)\,\phi_M(x,\mu) \,.
\end{equation}
The Gegenbauer moments have a diagonal scale evolution at leading order in perturbation theory. They are non-perturbative hadronic parameters, which can only be accessed using data or a non-perturbative approach such as light-cone QCD sum rules (see e.g.\ \cite{Ali:1993vd,Ball:1996tb,Ball:1998sk}) or lattice QCD \cite{Arthur:2010xf}. In Table~\ref{tab:hadronic_inputs} we collect the values for the decay constants and the first two Gegenbauer moments $a_{1,2}^M$ for light pseudoscalar and vector mesons. Our notation is such that $K^{(*)}\sim(q\bar s)$ with $q=u,d$, and $x$ is the momentum fraction of the light quark $q$.

\begin{table}
\begin{center}
\begin{tabular}{|c|c|cc|}
\hline 
Meson $M$ & $f_M$~[MeV] & $a_1^M(\mu_0)$ & $a_2^M(\mu_0)$ \\
\hline 
$\pi$ & $130.4\pm 0.2$ & 0 & $0.29\pm 0.08$ \\ 
$K$ & $156.2\pm 0.7$ & $-0.07\pm 0.04$ & $0.24\pm 0.08$ \\
$\rho$ & $212\pm 4$ & 0 & $0.17\pm 0.07$ \\
$\omega$ & $185\pm 5$ & 0 & $0.15\pm 0.12$ \\
$K^*$ & $203\pm 6$ & $-0.06\pm 0.04$ & $0.16\pm 0.09$ \\
$\phi$ & $231\pm 5$ & 0 & $0.23\pm 0.08$ \\
\hline 
\end{tabular}
\parbox{15.5cm}
{\caption{\label{tab:hadronic_inputs} 
Hadronic input parameters for light pseudoscalar and vector mesons, with scale-dependent quantities defined at $\mu_0=1$\,GeV. We assume isospin symmetry and use the same values for charged and neutral mesons. The values for $f_\pi$ and $f_K$ are taken from \cite{Agashe:2014kda}. The other decay constants are extracted from $\tau^-\to M^-\nu_\tau$ and $V^0\to l^+ l^-$ decays \cite{Neubert:1997uc}, as discussed in Appendix~\ref{app:decay_constants}. For all other parameters we adopt the values compiled in \cite{Dimou:2012un} from a combination of results obtained using lattice QCD \cite{Arthur:2010xf} and light-cone QCD sum rules (see e.g.\ \cite{Ball:1996tb,Ball:2005vx,Ball:2006wn,Ball:2006fz,Ball:2007rt}), including conservative error estimates.}}
\end{center}
\end{table} 

An expansion such as (\ref{Gegenbauer}) is useful provided we have some reason to believe that the infinite series is dominated by the first few terms. Higher-order Gegenbauer moments of the pion were studied in \cite{Bakulev:2001pa,Bakulev:2004cu} using a QCD sum-rule approach employing non-local vacuum condensates. These authors find $a_2^\pi=0.20$, $a_4^\pi=-0.14$, $a_6^\pi=5\cdot 10^{-3}$, and $a_8^\pi=a_{10}^\pi=4\cdot 10^{-3}$ at the scale $\mu_0=1$\,GeV. Their value of $a_2^\pi$ is consistent with the result given in Table~\ref{tab:hadronic_inputs}, while higher moments $a_n^\pi$ with $n\ge 6$ are estimated to be negligibly small. On the other hand, in more recent work \cite{Agaev:2012tm} the authors have performed fits to the first eight Gegenbauer moments of the pion LCDA using data on the $\pi^0\gamma^*\gamma$ form factor obtained by the BaBar and Belle collaborations \cite{Aubert:2009mc,Uehara:2012ag}. They find $a_2^\pi=0.10\,(0.14)$, $a_4^\pi=0.10\,(0.23)$, $a_6^\pi=0.10\,(0.18)$ and $a_8^\pi=0.034\,(0.050)$ at $\mu_0=1$\,GeV for Belle (BaBar), which suggests that $a_6^\pi$ and $a_8^\pi$ may not be insignificant. In our phenomenological analysis we will vary $a_4^M(\mu_0)$ between $-0.15$ and $+0.15$ for all light mesons and use this to estimate the effect of unknown higher Gegenbauer moments. With this treatment, the relevant combination of Gegenbauer coefficients given in relation (\ref{eq35}) below agrees with all of the above models within our quoted uncertainties.

It is an important question to ask what can be said on general grounds about the behavior of the Gegenbauer expansion. It is commonly assumed, and is supported by power-counting analyses in SCET, that the leading-twist LCDAs vanish at the endpoints $x=0$ and $x=1$, such that the integrals $\int_0^1\!\frac{dx}{x}\,\phi_M(x)$ and $\int_0^1\!\frac{dx}{1-x}\,\phi_M(x)$ converge. This statement implies that the infinite sums $\sum_n a_n^M$ and $\sum_n (-1)^n\,a_n^M$ converge. Barring accidental cancellations, this requires that for large $n$ the coefficients $a_n^M$ fall off faster than $1/n$, and this condition should hold for all values of $\mu_0$. From a physical point of view, high-rank Gegenbauer polynomials $C_n^{(3/2)}(2x-1)$ with $n\gg 1$ resolve structures on scales $\Delta x\sim 1/n$. For a light meson $M$, it is reasonable to assume that the LCDA $\phi_M(x)$ does not exhibit pronounced structures at scales much smaller than ${\cal O}(1)$, in which case the coefficients $a_n^M$ must decrease rapidly at large $n$. 

The LCDAs of heavy mesons are an exception to this rule, since the presence of the heavy-quark mass introduces a distinct scale. For a quarkonium state $M\sim (Q\bar Q)$ composed of two identical heavy quarks, the LCDA peaks at $x=1/2$ and has a width that tends to zero in the limit of infinite heavy-quark mass. The second moment of the LCDA around $x=1/2$ can be related to a local matrix element in non-relativistic quantum chromo-dynamics (NRQCD), the effective field theory describing heavy quarkonia states \cite{Caswell:1985ui,Bodwin:1994jh}. This framework provides a systematic expansion of hadronic matrix elements in powers of the small velocity $v\sim\alpha_s(m_Q v)$ of the heavy quark in the quarkonium rest frame. One obtains \cite{Braguta:2006wr}
\begin{equation}\label{mom2}
   \int_0^1\!dx \left( 2x - 1 \right)^2 \phi_M(x,\mu_0)
   = \frac{\langle v^2\rangle_M}{3} + {\cal O}(v^4) \,.
\end{equation}
To derive this result one uses that in the heavy-quark limit $x=\frac{p_Q\cdot\bar n}{2m_Q V\cdot\bar n}=\frac{1+v_z}{2}$, where $\bar n^\mu$ is a light-like vector, and $p_Q^\mu=m_Q V^\mu+k^\mu$ denotes the momentum of the heavy quark inside the quarkonium state with velocity $V^\mu$. The various vectors are defined such that $V\cdot\bar n=1$ and $V\cdot k=0$. In the rest frame of the quarkonium state we can choose $V^\mu=(1,\bm{0})$, $\bar n^\mu=(1,-\bm{e}_z)$, and $k^\mu=(0,m_Q\bm{v})$, where the 3-vector $\bm{v}$ is the residual velocity of the heavy quark inside the $(Q\bar Q)$ bound state. The factor 1/3 on the right-hand side of (\ref{mom2}) is due to rotational invariance in the rest frame. Numerical values for the NRQCD matrix element $\langle v^2\rangle$ for the $J/\psi$ and $\Upsilon(1S)$ states have been obtained from an analysis of the leptonic decay rates $\Gamma(J/\psi\to e^+ e^-)$ and $\Gamma(\Upsilon(1S)\to e^+ e^-)$ including first-order $\alpha_s$ corrections and non-perturbative contributions proportional to $v^2$. In this way, the values $\langle v^2\rangle_{J/\psi}=0.225\,_{-0.088}^{+0.106}$ \cite{Bodwin:2007fz} and $\langle v^2\rangle_{\Upsilon(1S)}=-0.009\pm 0.003$ \cite{Chung:2010vz} have been extracted, the latter one being inconsistent with the fact that the second moment in (\ref{mom2}) must be positive. Both estimates suffer from the fact that the two-loop \cite{Czarnecki:1997vz,Beneke:1997jm} and three-loop \cite{Beneke:2014qea} perturbative corrections to the NRQCD predictions for these decay rates are known to be huge, precluding a reliable extraction of non-perturbative parameters. Based on the power-counting rules of NRQCD one would naively expect that $\langle v^2\rangle_{J/\psi}\sim 0.3$ and $\langle v^2\rangle_{\Upsilon(1S)}\sim 0.1$, and we will use these estimates, along with a 50\% relative error assigned to them, in our phenomenological analysis. For our calculations we need the first inverse moments of the LCDA with respect to $x$ or $(1-x)$. Expanding the inverse moments about $x=1/2$, it is immediate to derive the model-independent relation \cite{Bodwin:2014bpa}
\begin{equation}\label{eq16}
   \int_0^1\!dx\,\frac{\phi_M(x,\mu_0)}{x} 
   = \int_0^1\!dx\,\frac{\phi_M(x,\mu_0)}{1-x}
   = 2 \left[ 1 + \frac{\langle v^2\rangle_M}{3} + {\cal O}(v^4) \right] .
\end{equation}
As a reasonable model at the low scale $\mu_0=1$\,GeV we adopt the Gaussian ansatz
\begin{equation}\label{LCDAQQ}
   \phi_M(x,\mu_0) = N_\sigma\,\frac{4x(1-x)}{\sqrt{2\pi}\sigma}\,
    \exp\left[ - \frac{(x-\frac12)^2}{2\sigma^2} \right] ; \qquad
   \sigma^2 = \frac{\langle v^2\rangle_M}{12} \,,
\end{equation}
where the polynomial in front of the Gaussian factor ensures that the LCDA vanishes at the endpoints $x=0,1$. The normalization constant $N_\sigma\approx 1$ can be expressed in closed form in terms of an error function. 

For a heavy-light meson state $M\sim (q\bar Q)$ composed of a light quark and a heavy anti-quark, the LCDA peaks at a small value $x\sim\Lambda_{\rm QCD}/m_M$, where $x$ refers to the momentum fraction of the light spectator quark. The appropriate effective field theory for heavy-light bound states is called heavy-quark effective theory (HQET), see \cite{Neubert:1993mb} for a review. In the context of this theory, it is possible to show that the first moment of the LCDA is determined by the ratio $\bar\Lambda_M/m_M$, where $m_M$ denotes the heavy-meson mass and $\bar\Lambda_M=m_M-m_Q$ (with $m_Q$ being the pole mass of the heavy quark) is a hadronic parameter. One obtains $\langle x\rangle=\frac{4}{3}\,\bar\Lambda_M/m_M+{\cal O}[\alpha_s(m_Q)]$ \cite{Grozin:1996pq}, where the one-loop radiative corrections have been calculated in \cite{Lee:2005gza} and are numerically significant. In our analysis below we need the first inverse moment of the LCDA with respect to $x$, which is of order $m_M/\Lambda_{\rm QCD}$ and cannot be related to a local HQET matrix element. One defines \cite{Beneke:1999br}
\begin{equation}\label{eq18}
   \int_0^1\!dx\,\frac{\phi_M(x,\mu_0)}{x}\equiv \frac{m_M}{\lambda_M(\mu_0)} + \dots \,, 
\end{equation}
where the hadronic parameter $\lambda_M(\mu_0)\sim\Lambda_{\rm QCD}$ is independent of the heavy-quark mass, and the dots denote corrections that are power-suppressed relative to the leading term. The parameter  $\lambda_M$ is poorly known at present. A QCD sum-rule estimate for the $B$ meson yields $\lambda_B(1\,{\rm GeV})=(460\pm 110)$\,MeV \cite{Braun:2003wx}, and we will use this value in our phenomenological analysis for both $B$ and $D$ mesons. Concerning $B_s$ and $D_s$ mesons, we shall use the estimate $\lambda_{M_s}-\lambda_M\approx 90$\,MeV from \cite{Ball:2006eu} and increase the error to $\pm 150$\,MeV. As a plausible model at a low scale $\mu_0=1$\,GeV we take \cite{Grozin:1996pq}
\begin{equation}\label{LCDAqQ}
   \phi_M(x,\mu_0) = N_\sigma\,\frac{x(1-x)}{\sigma^2}\,
    \exp\left( -\frac{x}{\sigma} \right) ; \qquad
   \sigma = \frac{\lambda_M(\mu_0)}{m_M} \,,
\end{equation}
where the normalization constant $N_\sigma\approx 1$ can be determined in closed form. For heavy-light mesons $M\sim (Q\bar q)$ containing a heavy quark and a light anti-quark, one simply replaces $x\leftrightarrow(1-x)$ in the above relations.

\begin{table}
\begin{center}
\begin{tabular}{|c|c|ccc|}
\hline 
Meson $M$ & $f_M$~[MeV] & $\lambda_M$~[MeV] & $\langle v^2\rangle$ & $\sigma$ \\
\hline 
$D$ & $204.6\pm 5.0$ & $460\pm 110$ & -- & $0.246\pm 0.059$ \\ 
$D_s$ & $257.5\pm 4.6$ & $550\pm 150$ & -- & $0.279\pm 0.076$ \\ 
$B$ & $186\pm 9$ & $460\pm 110$ & -- & $0.087\pm 0.021$ \\ 
$B_s$ & $224\pm 10$ & $550\pm 150$ & -- & $0.102\pm 0.028$ \\ 
$J/\psi$ & $403\pm 5$ & -- & $0.30\pm 0.15$ & $0.158\pm 0.040$ \\ 
$\Upsilon(1S)$ & $684\pm 5$ & -- & $0.10\pm 0.05$ & $0.091\pm 0.023$ \\ 
$\Upsilon(4S)$ & $326\pm 17$ & -- & $0.10\pm 0.05$ & $0.091\pm 0.023$ \\ 
\hline 
\end{tabular}
\parbox{15.5cm}
{\caption{\label{tab:hadronic_inputs2} 
Hadronic input parameters for pseudoscalar and vector mesons containing heavy quarks. Scale-dependent quantities are defined at $\mu_0=1$\,GeV. The values for $f_D$ and $f_{D_s}$ are taken from \cite{Agashe:2014kda}. The values for $f_B$ and $f_B$ are taken from two recent, unquenched lattice calculations \cite{Dowdall:2013tga,Bernardoni:2014fva}, which obtain identical central values but quote very different error estimates. We quote the averages of the uncertainties given by the two groups. The values of the $J/\psi$ and $\Upsilon(nS)$ decay constants can be derived from data, as explained in Appendix~\ref{app:decay_constants}.}}
\end{center}
\end{table} 

In Table~\ref{tab:hadronic_inputs2} we collect the values for the decay constants and the width parameters for heavy pseudoscalar and vector mesons, which will be used in our phenomenological analysis. In the cases of $(q\bar Q)$ and $(Q\bar Q)$ bound states, Gegenbauer moments of roughly $n\lesssim 1/\sigma$ give important contributions to the LCDAs, because they are required to resolve the narrow structures of the LCDAs near the peak region. For example, at $\mu_0=1$\,GeV the first 5 (6) Gegenbauer coefficients of the $B$-meson ($\Upsilon$-meson) LCDA are larger in magnitude than 0.1, and the first 7 (12) Gegenbauer coefficients are larger than 0.01. We will discuss in the next section that the effects of QCD evolution from a low scale $\mu_0$ up to a high scale reduces the high-rank Gegenbauer moments much stronger than Gegenbauer moments of low rank. For example, at $\mu=m_Z$ only the first 3 (2) Gegenbauer coefficients of the $B$-meson ($\Upsilon$-meson) LCDA are larger than 0.1, and the first 6 (8) Gegenbauer coefficients are larger than 0.01. As a result, the shapes of the LCDAs for mesons containing heavy quarks are significantly affected by RG evolution. For the case of the $B$-meson LCDA, this effect was studied in \cite{Lange:2003ff}. Consequently, the low-scale predictions for the inverse moments considered here are strongly modified at $\mu={\cal O}(m_Z)$.

\subsection{Radiative corrections and RG evolution}
\label{subsec:renorm}

\begin{figure}
\begin{center}
\includegraphics[width=0.2\textwidth]{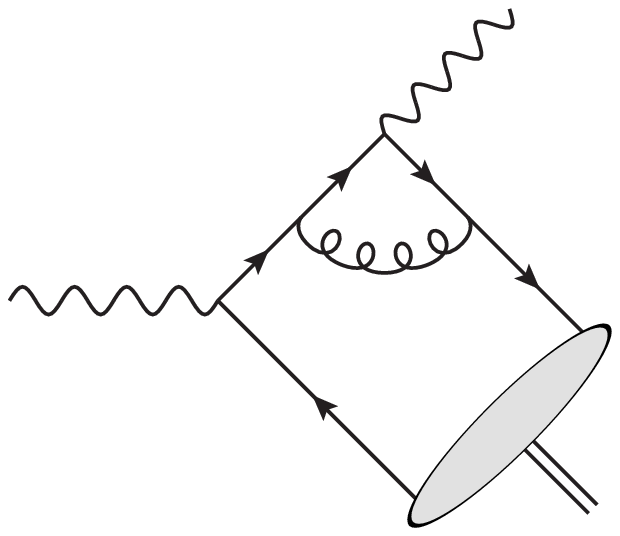}
\includegraphics[width=0.2\textwidth]{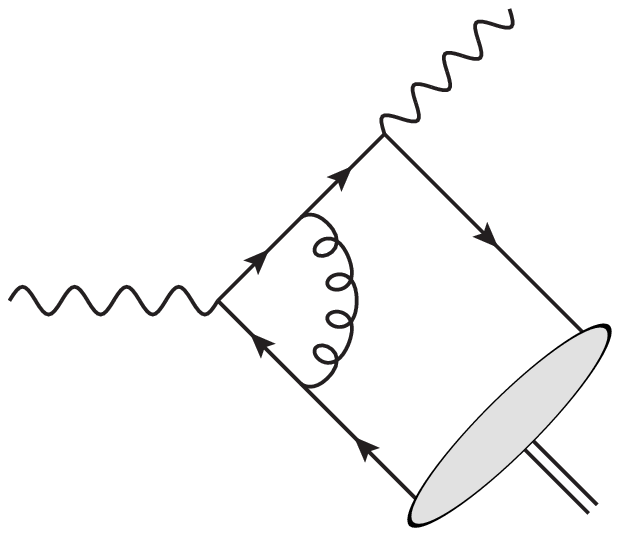}
\includegraphics[width=0.2\textwidth]{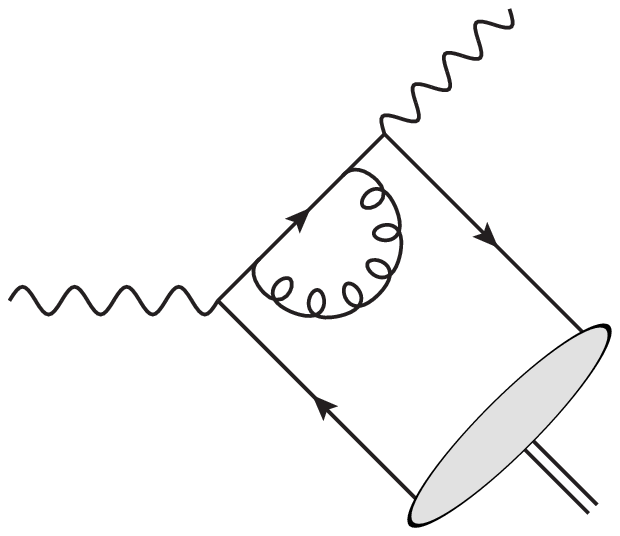}
\includegraphics[width=0.2\textwidth]{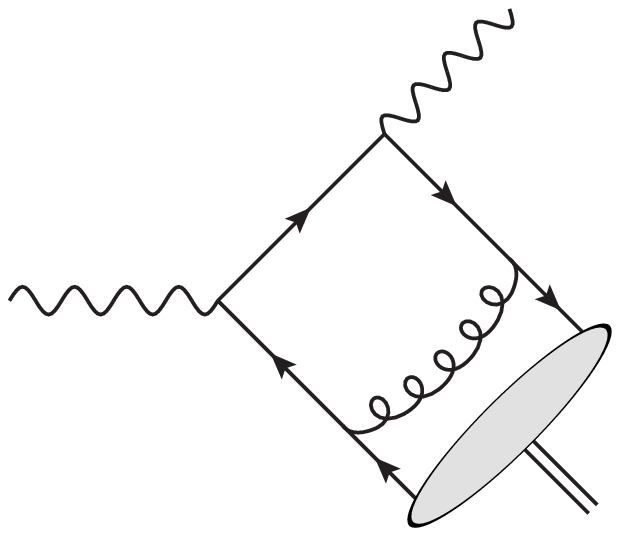}
\parbox{15.5cm}
{\caption{\label{fig:NLOgraphs}
One-loop QCD corrections to the first diagram in Figure~\ref{fig:LOgraphs}. Analogous corrections exist for the other diagram.}}
\end{center}
\end{figure}

In order to improve the accuracy of our predictions and be in a position to meaningfully discuss the setting of the factorization scale $\mu$, we include the ${\cal O}(\alpha_s)$ radiative corrections to the leading-twist contributions in our analysis, finding that RG evolution effects are very important. The reason is that logarithms of the form $\big(\alpha_s\ln(m_Z^2/\mu_0^2)\big)^n$, where $\mu_0\sim 1$\,GeV denotes the scale at which non-perturbative calculations of the LCDAs are performed, are numerically large and must be resummed to all orders of perturbation theory. We perform the calculation of the loop diagrams shown in Figure~\ref{fig:NLOgraphs} using dimensional regularization with $d=4-2\epsilon$ space-time dimensions. The individual on-shell graphs contain both UV and IR divergences. For the decays $Z\to M\gamma$ and $W\to M\gamma$, which are mediated by vector and axial-vector currents, the UV divergences cancel in the sum of all diagrams. The remaining IR poles are cancelled when we renormalize the LCDAs. To this end, we express the bare LCDAs in terms of the renormalized ones,
\begin{equation}
   \phi_M^{\rm bare}(x) = \int_0^1\!dy\,Z_\phi^{-1}(x,y,\mu)\,\phi_M(y,\mu) \,.
\end{equation}
At one-loop order, one obtains \cite{Lepage:1979zb,Chernyak:1983ej}
\begin{equation}
   Z_\phi(x,y,\mu) = \delta(x-y) + \frac{C_F\alpha_s(\mu)}{2\pi\epsilon}\,V_0(x,y) 
    + {\cal O}(\alpha_s^2) \,,
\end{equation}
where $C_F=(N_c^2-1)/(2N_c)=4/3$, and
\begin{equation}\label{BLkernel}
\begin{aligned}
   V_0(x,y) &= \frac12\,\delta(x-y) - \frac{1}{y(1-y)} \left[ x(1-y)\,\frac{\theta(y-x)}{y-x}
    + y(1-x)\,\frac{\theta(x-y)}{x-y} \right]_+ \\
   &\quad\mbox{}- \left[ \frac{x}{y}\,\theta(y-x) + \frac{1-x}{1-y}\,\theta(x-y) \right]
\end{aligned}
\end{equation}
is the one-loop Brodsky-Lepage kernel. For symmetric functions $g(x,y)$, the plus distribution is defined to act on test functions $f(x)$ as
\begin{equation}
   \int dy\,\big[ g(x,y) \big]_+\,f(x) = \int dy\,g(x,y)\,\big[ f(x) - f(y) \big] \,.
\end{equation}

Besides the subtraction of $1/\epsilon$ poles using dimensional regularization in the $\overline{\rm MS}$ scheme, one must carefully address the question of how to define $\gamma_5$ in $d\ne 4$ dimensions. Some of the amplitudes considered in this work involve traces of Dirac matrices containing a single insertion of $\gamma_5$. It is well known that for such traces the naive dimensional regularization scheme with anti-commuting $\gamma_5$ is algebraically inconsistent. Here we employ the 't\,Hooft-Veltman (HV) scheme \cite{'tHooft:1972fi}, in which $\gamma_5=i\gamma^0\gamma^1\gamma^2\gamma^3$ anti-commutes with the four matrices $\gamma^\mu$ with $\mu\in\{0,1,2,3\}$, while it commutes with the remaining $(d-4)$ Dirac matrices $\gamma_\perp^\mu$.\footnote{For the purposes of our analysis, the HV scheme is equivalent but more convenient than the scheme proposed by Larin \cite{Larin:1993tq}.}
While this definition is mathematically consistent, it violates the Ward identities of chiral gauge theories by finite terms, which must be restored order by order in perturbation theory \cite{Bonneau:1980ya}. In the present case, this is accomplished by performing the finite renormalization $A^\mu=Z_{\rm HV} A_{\rm HV}^\mu$ of the axial-vector current, where \cite{Trueman:1979en}
\begin{equation}\label{ZHV}
   Z_{\rm HV}(\mu) = 1 - \frac{C_F\alpha_s(\mu)}{\pi} + {\cal O}(\alpha_s^2) \,.
\end{equation}
In addition, the leading-twist LCDA of a pseudoscalar meson, which is defined in terms of a matrix element of a non-local axial-vector current on the light-cone, receives a finite renormalization of the form
\begin{equation}\label{phiPsubtr}
   \phi_{P,\,{\rm HV}}(x,\mu) = \int_0^1\!dy\,Z_{\rm HV}^{-1}(x,y,\mu)\,\phi_P(y,\mu) \,,
\end{equation}
where \cite{Melic:2001wb}
\begin{equation}
   Z_{\rm HV}^{-1}(x,y,\mu) = \delta(x-y) + \frac{2C_F\alpha_s(\mu)}{\pi} 
    \left[ \frac{x}{y}\,\theta(y-x) + \frac{1-x}{1-y}\,\theta(x-y) \right]
    + {\cal O}(\alpha_s^2) \,.
\end{equation}
This redefinition is important to restore the proper normalization of the LCDA $\phi_P(x,\mu)$. Integrating relation (\ref{phiPsubtr}) over $x$, we find that
\begin{equation}
   \int_0^1\!dx\,\phi_{P,\,{\rm HV}}(x,\mu) 
   = Z_{\rm HV}^{-1}(\mu)\,\int_0^1\!dy\,\phi_P(y,\mu) = Z_{\rm HV}^{-1}(\mu) \,,
\end{equation}
with $Z_{\rm HV}$ given in (\ref{ZHV}). The integral turns the matrix element of the non-local axial-vector current into the corresponding local matrix element.

Our final expressions for the decay amplitudes will contain the scale-dependent, leading-twist LCDAs $\phi_M(x,\mu)$ with $M=P,V_\parallel$. These functions satisfy the integro-differential evolution equation
\begin{equation}
   \mu\,\frac{d}{d\mu}\,\phi_M(x,\mu) = - \int_0^1\!dy\,V(x,y,\mu)\,\phi_M(y,\mu) \,,
\end{equation}
where $V(x,y,\mu)=V_0(x,y)\,\frac{C_F\alpha_s(\mu)}{\pi}+{\cal O}(\alpha_s^2)$. The eigenfunctions of the one-loop Brodsky-Lepage kernel $V_0(x,y)$ in (\ref{BLkernel}) are the Gegenbauer polynomials $6x(1-x)\,C_n^{(3/2)}(2x-1)$, and hence the Gegenbauer moments $a_n(\mu)$ defined in (\ref{Gegenbauer}) are multiplicatively renormalized at this order. They obey the RG equation \cite{Lepage:1979zb}
\begin{equation}\label{RGEs}
   \mu\,\frac{d}{d\mu}\,a_n^M(\mu) = - \gamma_n\,\frac{\alpha_s(\mu)}{4\pi}\,a_n^M(\mu) \,, 
\end{equation}
where
\begin{equation}\label{gamma_n}
   \gamma_n = 2C_F \left( 4 H_{n+1} - \frac{2}{(n+1)(n+2)} - 3 \right) , \qquad
   \mbox{with} \quad H_{n+1} = \sum_{k=1}^{n+1}\,\frac{1}{k} \,.
\end{equation}
The evolution of the leading-twist LCDAs at two-loop order has been studied in \cite{Dittes:1983dy,Mikhailov:1984ii,Mueller:1993hg,Mueller:1994cn}. The RG equation for the Gegenbauer moments becomes more complicated at this order, since the scale dependence of $a_n^M(\mu)$ receives contributions proportional to $a_k^M(\mu)$ with $k=0,\dots,n$ \cite{Brodsky:1984xk,Mueller:1993hg,Mueller:1994cn}. The evolution equation can still be solved analytically using an iterative scheme. Explicit results for the moments up to $n=12$ can be found in \cite{Agaev:2010aq}. However, given that all present estimates of the hadronic parameters $a_n^M$ are afflicted with large theoretical uncertainties, it is sufficient for all practical purposes to use the leading-order solution (\ref{RGEs}). It reads
\begin{equation}\label{anevol}
   a_n^M(\mu) = \left( \frac{\alpha_s(\mu)}{\alpha_s(\mu_0)} \right)^{\gamma_n/2\beta_0} 
    a_n^M(\mu_0) \,,
\end{equation}
where $\beta_0=\frac{11}{3} N_c-\frac23 n_f$ is the first coefficient of the QCD $\beta$ function. Here $\mu_0\sim 1$\,GeV denotes a low scale, at which the Gegenbauer moments are derived from a non-perturbative approach, while $\mu$ is a high scale to which the LCDAs are evolved. In our analysis this scale is set by the mass of the decaying electroweak boson. Note that one must adjust the values of $\beta_0$ whenever $\mu$ crosses a flavor threshold. All of the anomalous dimensions are strictly positive, which implies that $a_n^M(\mu)\to 0$ in the formal limit $\mu\to\infty$. Indeed, for large $n$ the evolution supplies an additional suppression factor $(1/n)^K$ with $K=\frac{C_F\alpha_s}{\pi}\ln\frac{\mu^2}{\mu_0^2}$. In this limit, the leading-twist LCDAs approach the asymptotic form $6x(1-x)$. 

\begin{figure} 
\begin{center}
\includegraphics[height=0.262\textwidth]{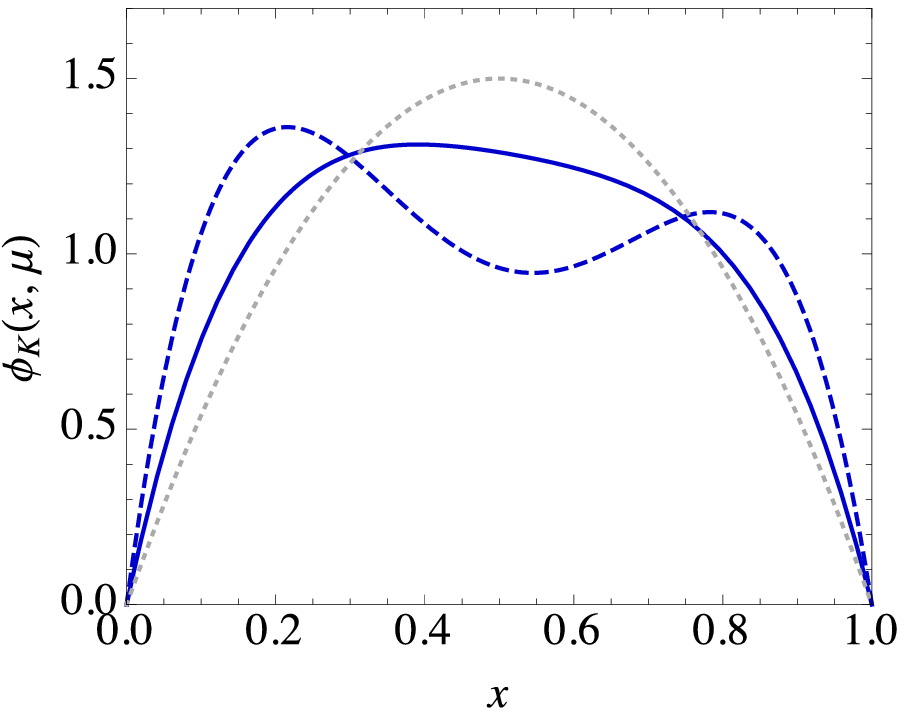}
\includegraphics[height=0.268\textwidth]{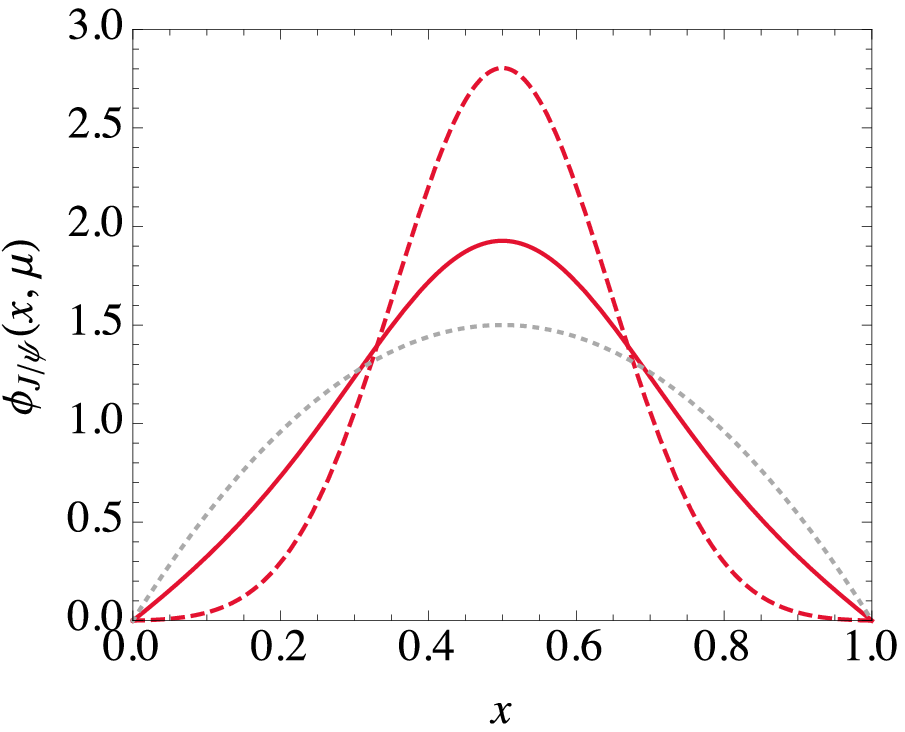}
\includegraphics[height=0.268\textwidth]{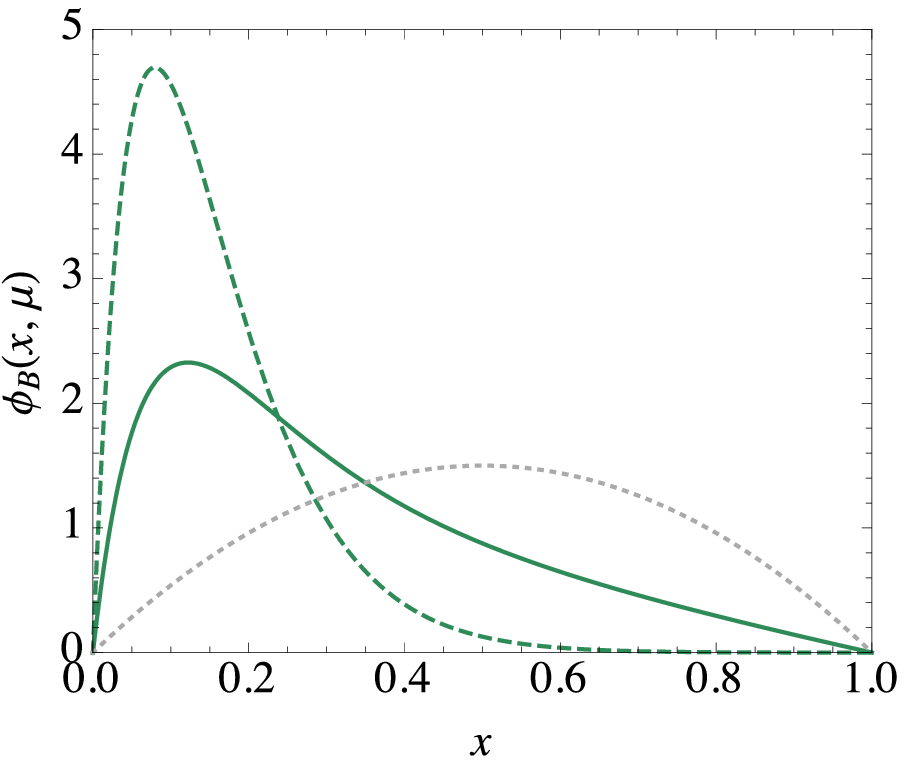}
\parbox{15.5cm}
{\caption{\label{fig:LCDAevol} 
RG evolution of the LCDAs of the kaon (left), the $J/\psi$ meson (middle) and the $B$ meson (right) from a low scale $\mu_0=1$\,GeV (dashed lines) to a high scale $\mu=m_Z$ (solid lines). The dotted grey line shows the asymptotic form $6x(1-x)$ for comparison.}}
\end{center}
\end{figure}

Figure~\ref{fig:LCDAevol} shows the RG evolution of the LCDAs of the kaon, $J/\psi$ meson and $B$ meson from a low scale $\mu_0=1$\,GeV up to a high scale $m_Z$. We use the Gegenbauer moments and width parameters collected in Tables~\ref{tab:hadronic_inputs} and \ref{tab:hadronic_inputs2}. For light mesons we truncate the Gegenbauer expansion (\ref{Gegenbauer}) at $n=2$. For heavy mesons we use the model LCDAs given in (\ref{LCDAQQ}) and (\ref{LCDAqQ}), compute their first 20 Gegenbauer moments, evolve the corresponding coefficients $a_n^M$ from $\mu_0$ to $m_Z$, and reconstruct the LCDAs at the high scale from (\ref{Gegenbauer}). The dotted line in the plots shows the asymptotic form $6x(1-x)$. Evolution effects alter the shapes of the various distributions in a significant way. At the electroweak scale, the LCDAs are significantly closer to the asymptotic form $6x(1-x)$ than at a low hadronic scale. Consequently, RG effects render our predictions more insensitive to poorly determined hadronic input parameters. Notice, in particular, that the LCDA of the $J/\psi$ meson at $\mu=m_Z$ is as close to the asymptotic form as the kaon LCDA. In practice, the LCDAs of heavy mesons at a scale much larger than the heavy-quark mass can be well described in terms of a Gegenbauer expansion truncated after a few Gegenbauer moments.

\subsection{Flavor wave functions of neutral mesons}

The couplings of photons and of the electroweak gauge bosons $W$ and $Z$ to fermions are flavor dependent. While the flavor content of charged mesons is unambiguous, for neutral mesons complications arise from the fact that a given meson can be a superposition of different flavor components. We write the flavor wave function of the neutral final-state meson $M^0$ in the form
\begin{equation}
   |M^0\rangle = \sum_{q=u,d,s,c,b}\,c_q^M\,|q\bar q\rangle \,; 
    \quad \mbox{with} \quad \sum_q |c_q^M|^2 = 1 \,.
\end{equation}
For heavy mesons containing charm or bottom quarks such effects can safely be neglected. The heavy mesons $\eta_c$ and $J/\psi$ have $c_c=1$, while $\eta_b$ and $\Upsilon$ have $c_b=1$. Mixing effects can however be important for light mesons.

Following \cite{Beneke:2002jn}, we assume isospin symmetry of all hadronic matrix elements, but we differentiate between the matrix elements of mesons containing up or down quarks and those containing strange quarks. The $\pi^0$ and $\rho^0$ mesons are members of an isospin triplet and have flavor content $(|u\bar u\rangle-|d\bar d\rangle)/\sqrt2$. Things get more complicated when we consider the mesons $\eta$, $\eta'$ and $\omega$, $\phi$, however. In the $SU(3)$ flavor-symmetry limit, the pseudoscalar meson $\eta$ is a flavor octet and $\eta'$ a flavor singlet. However, it is known empirically that $SU(3)$-breaking corrections to these assignments are large. In the following we shall not rely on $SU(3)$ flavor symmetry, but instead introduce another assumption, expected to be accurate at the 10\% level. In the absence of the axial anomaly, the flavor states $|\eta_q\rangle=(|u\bar u\rangle+|d\bar d\rangle)/\sqrt2$ and $\eta_s\rangle=|s\bar s\rangle$ mix only through OZI-violating effects, which are known phenomenologically to be small. It is therefore reasonable to assume that the axial anomaly is the only effect that mixes the two flavor states \cite{Feldmann:1998vh,Feldmann:1999uf}. This assumption implies, in particular, that the vector mesons $\omega$ and $\phi$ are pure $(|u\bar u\rangle+|d\bar d\rangle)/\sqrt2$ and $|s\bar s\rangle$ states, respectively, as is indeed the case to very good approximation. The anomaly introduces an effective mass term for the system of $\eta$ and $\eta'$ states, which is not diagonal in the flavor basis $\{|\eta_q\rangle,\,|\eta_s\rangle\}$. Since this is by assumption the only mixing effect, one obtains a mixing scheme with a single mixing angle in the flavor basis. 

As explained in \cite{Beneke:2002jn}, the $\eta$ and $\eta'$ mesons have a leading-twist two-gluon LCDA besides the LCDAs corresponding to the quark-anti-quark Fock states $\eta_q$ and $\eta_s$. The two-gluon LCDA contributes to the $Z\to\eta^{(\prime)}\gamma$ decay amplitudes at order $\alpha_s$, through fermion box graphs with $Z\gamma gg$ as external particles. A detailed analysis of these decays will be presented elsewhere~\cite{inprep}.

\section{Radiative decays of electroweak gauge bosons}
\label{sec:raddecays}

We now apply our general approach to study the rare, exclusive radiative decays $Z\to M\gamma$ and $W\to M\gamma$, where $M$ denotes a pseudoscalar ($P$) or vector meson ($V$). The leading-order Feynman diagrams contributing to the first process were already shown in Figure~\ref{fig:LOgraphs}. We only consider cases where the mass of the final-state meson is much smaller than the mass of the decaying boson. Up to corrections of order $(m_M/m_{Z,W})^2$ this mass can then be set to zero. 

\subsection{\boldmath Radiative hadronic decays of $Z$ bosons}

We begin our analysis with the decays $Z^0\to M^0\gamma$. We find that, at leading order in the expansion in $\Lambda_{\rm QCD}/m_Z$, only pseudoscalar or longitudinally polarized vector mesons can be produced. The corresponding decay amplitudes can be written in the general form
\begin{equation}\label{ampl1}
   i{\cal A}(Z\to M\gamma)
   = \pm\frac{eg f_M}{2\cos\theta_W} \left[ i\epsilon_{\mu\nu\alpha\beta}\,
    \frac{k^\mu q^\nu\varepsilon_Z^\alpha\,\varepsilon_\gamma^{*\beta}}{k\cdot q}\,F_1^M 
    - \left( \varepsilon_Z\cdot\varepsilon_\gamma^* 
    - \frac{q\cdot\varepsilon_Z\,k\cdot\varepsilon_\gamma^*}{k\cdot q} \right) F_2^M 
    \right] ,
\end{equation}
where the upper (lower) sign refers to the case where $M=P$ ($V_\parallel$). Here $\theta_W$ is the electroweak mixing angle. Both the photon and the $Z$ boson are transversely polarized with respect to the decay axis. The second term inside the brackets can be written more compactly as $\varepsilon_Z^\perp\cdot\varepsilon_\gamma^{\perp*}$, and below we use this as a short-hand notation. We use a convention where $\epsilon_{0123}=-1$. For neutral mesons that are eigenstates of the charge-conjugation operation, $C$ invariance implies~\cite{Arnellos:1981gy}
\begin{equation}\label{Godd}
   F_2^M = 0 \,.
\end{equation}
The decay amplitudes are then proportional to the vector product $\bm{\varepsilon}_Z\times\bm{\varepsilon}_\gamma^*$ of the transversely polarized photon and $Z$ boson. However, in new-physics models in which the $Z$ boson has flavor-changing neutral-current (FCNC) couplings, $Z\to M\gamma$ decays into mesons that are not flavor diagonal (and hence not eigenstates of $C$) can occur. In this case relation (\ref{Godd}) no longer holds. In complete generality, the decay rates, summed (averaged) over the polarization states of the photon ($Z$ boson), are obtained as
\begin{equation}\label{Zrates}
   \Gamma(Z\to M\gamma) = \frac{\alpha m_Z f_M^2}{6v^2} 
    \left( \left| F_1^M \right|^2 + \left| F_2^M \right|^2 \right) .
\end{equation}
Here $\alpha=1/137.036$ is the fine-structure constant evaluated at $q^2=0$ \cite{Agashe:2014kda}, as appropriate for a real photon, and $v$ denotes the Higgs vacuum expectation value, which enters through the relation $(g/\cos\theta_W)^2=4m_Z^2/v^2$ evaluated at $\mu=m_Z$. This can be solved to give
\begin{equation}\label{v_value}
   v\equiv v(m_Z) = m_Z\,\frac{\sin\theta_W\cos\theta_W}{\sqrt{\pi\alpha(m_Z)}}
   = 245.36\,\mbox{GeV} \,,
\end{equation}
where we have used $\alpha(m_Z)=1/127.940\pm 0.014$ and $\sin^2\theta_W=0.23126\pm 0.00005$, with the weak mixing angle determined from the neutral-current couplings of the $Z$ boson evaluated at $\mu=m_Z$ \cite{Agashe:2014kda}. The form factors $F_i^M$ are given in terms of overlap integrals of calculable hard-scattering coefficients with LCDAs. 

Evaluating the diagrams shown in Figures~\ref{fig:LOgraphs} and \ref{fig:NLOgraphs}, we find that the relevant hard-scattering coefficients for the decays $V\to M\gamma$ (with $V=Z,W$) are given by 
\begin{equation}
   H_\pm(x,m_V,\mu) = \frac{1}{x} \left[ 1 + \frac{C_F\alpha_s(\mu)}{4\pi}\,h_\pm(x,m_V,\mu) 
   + {\cal O}(\alpha_s^2) \right] ,
\end{equation}
where
\begin{equation}\label{gpm}
   h_\pm(x,m_V,\mu) = (2\ln x+3) \left( \ln\frac{m_V^2}{\mu^2} - i\pi \right) 
    + \ln^2 x - 9 + (\pm 1-2)\,\frac{x\ln x}{1-x} \,.
\end{equation}
Our result for $h_+$ agrees with a corresponding expression derived in the context of a study of meson-photon transition form factors at high $Q^2$ performed in \cite{Braaten:1982yp}. The expression for $h_-$ is new. The relevant convolutions of the hard-scattering coefficients $H_\pm(x,m_V,\mu)$ with LCDAs give rise to the master integrals (we define $a_0^M(\mu)\equiv 1$)
\begin{equation}\label{Ipmdef}
\begin{aligned}
   I_\pm^M(m_V) &= \int_0^1\!dx\,H_\pm(x,m_V,\mu)\,\phi_M(x,\mu)
    = 3\,\sum_{n=0}^\infty\,(-1)^n\,C_n^{(\pm)}(m_V,\mu)\,a_n^M(\mu) \,, \\
   \bar I_\pm^M(m_V) &= \int_0^1\!dx\,H_\pm(1-x,m_V,\mu)\,\phi_M(x,\mu)
    = 3\,\sum_{n=0}^\infty\,C_n^{(\pm)}(m_V,\mu)\,a_n^M(\mu) \,,
\end{aligned}
\end{equation}
with
\begin{equation}\label{Cndef}
   C_n^{(\pm)}(m_V,\mu) = 1 + \frac{C_F\alpha_s(\mu)}{4\pi}\,c_n^{(\pm)}\Big(\frac{m_V}{\mu}\Big) 
    + {\cal O}(\alpha_s^2) \,.
\end{equation}
The integrals $I_\pm^M$ arise from the diagrams shown in Figure~\ref{fig:NLOgraphs}, in which the photon is attached to the quark inside the meson. Diagrams in which the photon is attached to the anti-quark give rise to the integrals $\bar I_\pm^M$. In evaluating the integrals we have used the Gegenbauer expansion (\ref{Gegenbauer}). The two types of integrals are related to each other by the fact that the Gegenbauer polynomials $C_n^{(3/2)}(2x-1)$ transform into themselves times a factor $(-1)^n$ under the exchange of $x\leftrightarrow(1-x)$. Notice that at tree level the master integrals involve the infinite sums over Gegenbauer moments with equal coefficients. Employing a technique explained in Appendix~\ref{app:moments}, we have succeeded to derive a closed expression for the one-loop coefficients $c_n^{(\pm)}(m_V/\mu)$. It reads
\begin{equation}\label{gorgeous}
\begin{aligned}
   c_n^{(\pm)}\Big(\frac{m_V}{\mu}\Big) &= \left[ \frac{2}{(n+1)(n+2)} - 4 H_{n+1} + 3 \right] 
    \left( \ln\frac{m_V^2}{\mu^2} - i\pi \right) \\
   &\quad\mbox{}+ 4 H_{n+1}^2 - \frac{4(H_{n+1}-1)\pm 1}{(n+1)(n+2)}
    + \frac{2}{(n+1)^2 (n+2)^2} - 9 \,.    
\end{aligned}
\end{equation}
Note that $c_0^{(+)}=-5$ and $c_0^{(-)}=-4$ are pure numbers. Using the evolution equations (\ref{RGEs}) and the explicit expressions for the one-loop anomalous dimensions given in (\ref{gamma_n}), it is straightforward to check that the master integrals in (\ref{Ipmdef}) are independent of the factorization scale $\mu$. Indeed, the coefficient of the logarithm in (\ref{gorgeous}) is equal to $-\gamma_n/(2C_F)$. Note also that the imaginary parts associated with the logarithm do not contribute to the decay rates at ${\cal O}(\alpha_s)$.

\begin{figure}[t] 
\begin{center}
\includegraphics[height=0.35\textwidth]{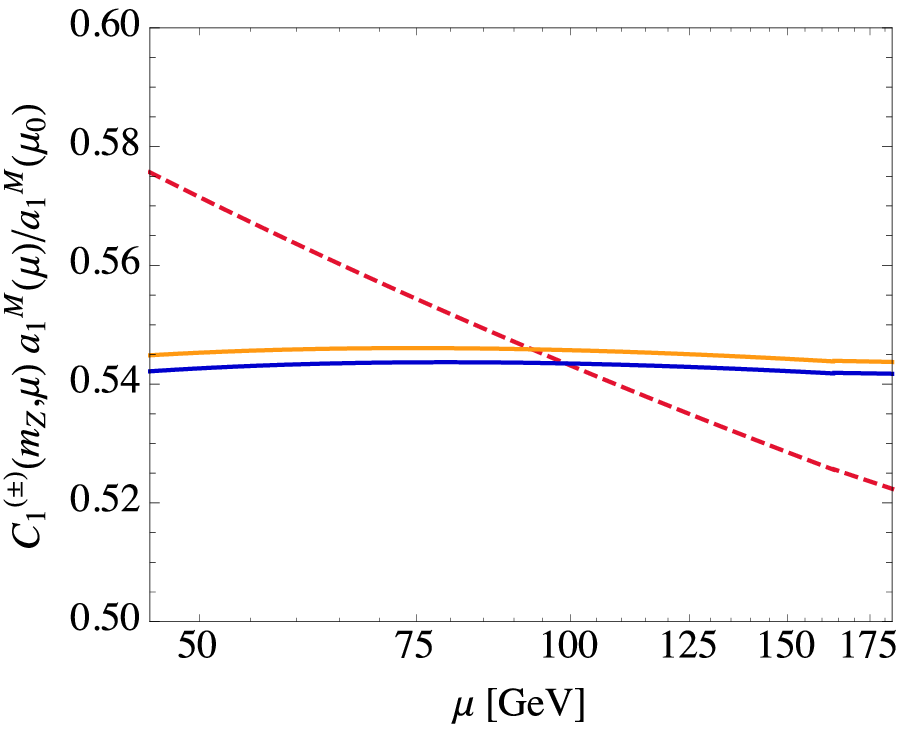} \qquad
\includegraphics[height=0.35\textwidth]{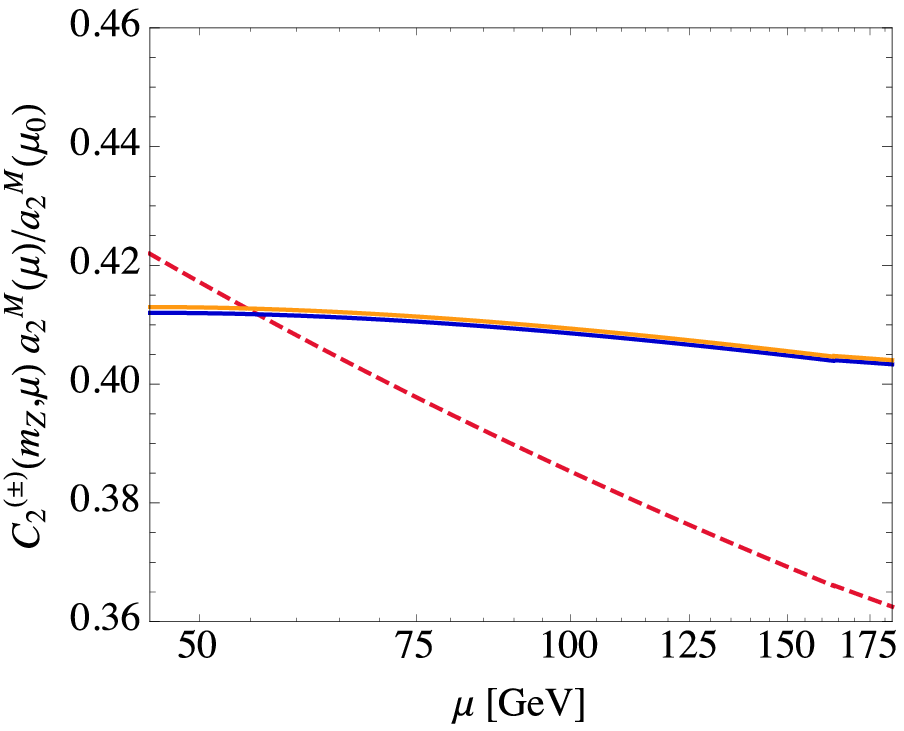} 
\parbox{15.5cm}
{\caption{\label{fig:scaledep} 
Scale dependence of the combinations $C_n^{(\pm)}(m_Z,\mu)\,a_n^M(\mu)/a_n^M(\mu_0)$ for the first two Gegenbauer moments ($n=1,2$). The red dashed lines show the results at leading-order, where $C_n^{(\pm)}(m_Z,\mu)=1$. The blue and yellow lines show the results at next-to-leading order obtained when the one-loop expressions in (\ref{Cndef}) are used.}}
\end{center}
\end{figure}

In Figure~\ref{fig:scaledep}, we study the scale dependence of individual terms in the sums over Gegenbauer moments in (\ref{Ipmdef}) at leading (dashed red lines) and next-to-leading order (solid lines) in perturbation theory. At leading order the $\mu$ dependence of the Gegenbauer moments, shown explicitly in (\ref{anevol}), is left uncompensated, and hence a significant scale dependence arises. At next-to-leading order this dependence is compensated by the logarithmic terms contained in the one-loop corrections (\ref{gorgeous}) to the hard-scattering coefficients $C_n^{(+)}(m_Z,\mu)$ (blue lines) and $C_n^{(-)}(m_Z,\mu)$ (orange lines). The resulting next-to-leading order curves exhibit excellent stability under variations of the factorization scale in the interval $m_Z/2<\mu<2m_Z$.

In terms of the master integrals defined in (\ref{Ipmdef}), the form factors $F_i^M$ are given by
\begin{equation}\label{FVPres}
\begin{aligned}
   F_1^M &= \frac{{\cal Q}_M}{6} \left[ I_+^M(m_Z) + \bar I_+^M(m_Z) \right] 
    = \phantom{-} {\cal Q}_M \sum_{n=0}^\infty\,C_{2n}^{(+)}(m_Z,\mu)\,a_{2n}^M(\mu) \,, \\
   F_2^M &= \frac{{\cal Q}_M'}{6} \left[ I_-^M(m_Z) - \bar I_-^M(m_Z) \right] 
    = - {\cal Q}_M' \sum_{n=0}^\infty\,C_{2n+1}^{(-)}(m_Z,\mu)\,a_{2n+1}^M(\mu) \,,
\end{aligned}
\end{equation}
where
\begin{equation}
   {\cal Q}_P = \sum_q\,6 c_q^P Q_q\,v_q \,, \qquad
    {\cal Q}_V = \sum_q\,6 c_q^V Q_q\,a_q \,, 
\end{equation}
and the coefficients ${\cal Q}_M'$ and related to ${\cal Q}_M$ by exchanging $v_q\leftrightarrow a_q$. Corrections to the results (\ref{FVPres}) arise only at twist-4 level and are suppressed by $(\Lambda_{\rm QCD}/m_Z)^2$ or $(m_M/m_Z)^2$. They are phenomenologically irrelevant. In the above expressions $Q_q$ denotes the electric charge of a quarks in units of $e$, while $v_q=\frac12\,T_3^q-\sin^2\theta_W\,Q_q$ and $a_q=\frac12\,T_3^q$ (not to be confused with the Gegenbauer moments) are its vector and axial-vector couplings to the $Z$ boson. Our finding that the form factors for pseudoscalar and vector mesons in (\ref{FVPres}) have exactly the same structure crucially relies on a mathematically consistent treatment of $\gamma_5$, see Section~\ref{subsec:renorm}. At tree level $C_n^{(\pm)}=1$, and hence the form factor $F_1^M$ ($F_2^M$) is proportional to the infinite sum of all even (odd) Gegenbauer moments of the meson $M$. Charge-conjugation invariance implies that the LCDAs of a flavor-diagonal neutral mesons are symmetric under the exchange of $x$ and $(1-x)$, and hence for these mesons the odd Gegenbauer moments $a_{2n+1}^M$ vanish. This leads to relation (\ref{Godd}). The non-zero form factor $F_1^M$ involves the infinite sum over the even Gegenbauer moments times some flavor-dependent coefficients ${\cal Q}_M$, which we collect in Table~\ref{tab:coefs}.

\begin{table}
\begin{center}
\begin{tabular}{|c|c||c|c|}
\hline 
Meson $P$ & ${\cal Q}_P=\sum_q 6 c_q^P Q_q\,v_q$ & Meson $V$ & ${\cal Q}_V=\sum_q 6 c_q^V Q_q\,a_q$ \\ 
\hline 
$\pi^0$ & $\frac{1}{2\sqrt2} \left( 1-4\sin^2\theta_W \right)$ & $\rho^0$ & $\frac{1}{2\sqrt2}$ \\ 
$\eta_q$ & $\frac{3}{2\sqrt2} \left( 1-\frac{20}{9} \sin^2\theta_W \right)$ & $\omega$ & $\frac{3}{2\sqrt2}$ \\ [1mm]
$\eta_s$, $\eta_b$ & $\frac12 - \frac23\sin^2\theta_W$ & $\phi$, $\Upsilon$ & $\frac12$ \\ [1mm]
$\eta_c$ & $1-\frac83\sin^2\theta_W$ & $J/\psi$ & 1 \\ [0.5mm]
\hline 
\end{tabular}
\parbox{15.5cm}
{\caption{\label{tab:coefs} 
Coefficients ${\cal Q}_M$ for the ground-state neutral pseudoscalar and vector mesons.}}
\end{center}
\end{table} 

Explicit predictions for the leading-twist LCDAs derived by means of non-perturbative methods are typically obtained at a low hadronic scale $\mu_0\sim 1$\,GeV. When these predictions are used in (\ref{FVPres}), the expressions for the radiative corrections involve large logarithms $\ln(m_Z^2/\mu_0^2)\approx 9$, which must be resummed to all orders in perturbation theory to obtain reliable predictions. This resummation is most readily performed by evaluating the result (\ref{FVPres}) at the scale $\mu=m_Z$ (or any other scale of the same order), in which case we obtain
\begin{equation}\label{eq35}
\begin{aligned}
   \mbox{Re}\,F_1^M &= {\cal Q}_M\,\Big[ 0.94 + 1.05\,a_2^M(m_Z) + 1.15\,a_4^M(m_Z) 
    + 1.22\,a_6^M(m_Z) + \dots \Big] \\
   &= {\cal Q}_M\,\Big[ 0.94 + 0.41\,a_2^M(\mu_0) + 0.29\,a_4^M(\mu_0) 
    + 0.23\,a_6^M(\mu_0) + \dots \Big] \,.
\end{aligned}
\end{equation}
We use the three-loop expression for the running coupling as provided by the {\tt RunDec} program \cite{Chetyrkin:2000yt}, normalized to $\alpha_s(m_Z)=0.1185\pm 0.0006$ \cite{Agashe:2014kda} and with heavy-quark thresholds at $m_b(m_b)=4.163$\,GeV and $m_c(m_c)=1.279$\,GeV \cite{Chetyrkin:2009fv}. The Gegenbauer moments at the high scale $\mu=m_Z$ in the first line can be related to hadronic input parameters calculated at the low scale $\mu_0=1$\,GeV using the relations (\ref{anevol}). In this process the coefficients of the higher moments get successively smaller.

Decays into a transversely polarized vector meson are only allowed at twist-3 order. This presents us with an opportunity to study the structure of power corrections with a specific test case. We adopt the approximation where three-particle LCDAs are neglected. We then evaluate the diagrams in Figure~\ref{fig:LOgraphs} using the projector for a transversely polarized vector meson given in Appendix~\ref{app:LCDAs}. The decay amplitude can be decomposed in a form analogous to (\ref{ampl1}), such that
\begin{equation}\label{amplperp}
   i{\cal A}(Z_\parallel\to V_\perp\gamma)
   = - \frac{eg f_V}{2\cos\theta_W}\,\frac{m_V}{m_Z} \left( i\epsilon_{\mu\nu\alpha\beta}\,
    \frac{k^\mu q^\nu\varepsilon_V^{*\alpha} \varepsilon_\gamma^{*\beta}}{k\cdot q} F_1^\perp 
    - \varepsilon_V^{\perp *}\cdot\varepsilon_\gamma^{\perp *}\,F_2^\perp \right) .
\end{equation}
The $Z$ boson must be longitudinally polarized, and its polarization vector can be written as $\varepsilon_Z^\mu=(q-k)^\mu/m_Z$. The extra factor of $m_V/m_Z$ compared with (\ref{ampl1}) makes the power suppression of these amplitudes explicit. The corresponding decay rate, summed (averaged) over the polarizations of the final-state (initial-state) particles, is given by
\begin{equation}
   \Gamma(Z\to V_\perp\gamma) = \frac{\alpha m_Z f_V^2}{6v^2}\,\frac{m_V^2}{m_Z^2}
    \left( \left| F_1^\perp \right|^2 + \left| F_2^\perp \right|^2 \right) .
\end{equation}
The general expressions for the form factors $F_i^\perp$ in terms of overlap integrals over the various twist-3 LCDAs appearing in the projector (\ref{LCDAVperp}) for a transversely polarized vector meson are given in Appendix~\ref{app:LCDAs}. They can be simplified a lot by using relations implied by the equations of motion in the limit where three-particle LCDAs are neglected. Assuming for simplicity that quark-mass effects can be neglected, we obtain
\begin{equation}\label{HAVperpres}
   F_1^\perp = - \frac{{\cal Q}_V}{3} \int_0^1\!dx 
    \left( \frac{\ln x}{1-x} + \frac{\ln(1-x)}{x} \right) \phi_V(x,\mu) 
    + \mbox{3-particle LCDAs} \,.
\end{equation}
An analogous expression with ${\cal Q}_V$ replaced by ${\cal Q}_V'$ and a relative minus sign between the two terms inside the parenthesis holds for $F_2^\perp$. Since the leading-twist LCDAs of neutral mesons are symmetric in $x\leftrightarrow (1-x)$, it follows that $F_2^\perp=0$, and hence once again only the term involving the Levi-Civita tensor in (\ref{amplperp}) contributes to the decay amplitude. Using the Gegenbauer expansion of the LCDA $\phi_V(x,\mu)$ in (\ref{Gegenbauer}), we obtain
\begin{equation}
   F_1^\perp = {\cal Q}_V \left[ 1 + \sum_{n=1}^\infty \frac{a_{2n}^V(\mu)}{(n+1)(2n+1)} \right] .
\end{equation}
Since we have not evaluated radiative corrections to the form factors, we do not control the scale dependence of the Gegenbauer moments $a_n^V$. However, it is clear that in order to avoid large logarithms we should again set $\mu\approx m_Z$ in the final result. Note that the result for $F_1^\perp$ is similar to that for $F_1^V$ in (\ref{FVPres}), but the coefficients of higher Gegenbauer moments are more strongly suppressed. For a rough estimate, we may assume that the Gegenbauer moments only have a minor impact on the final results (i.e.\ $a_n^V(m_Z)\gg 1$), in which case it follows that the rates for $Z\to V\gamma$ decays with transversely and longitudinally polarized vector mesons are related by
\begin{equation}\label{smallrat}
   \frac{\Gamma(Z\to V_\perp\gamma)}{\Gamma(Z\to V_\parallel\gamma)} 
   \approx \frac{m_V^2}{m_Z^2} \,.
\end{equation}
This ratio is of order $10^{-4}$. The outcome of this discussion is that power corrections in the expansion in $\Lambda_{\rm QCD}/m_Z$ are completely negligible for phenomenological applications.

\subsection{\boldmath Radiative hadronic decays of $W$ bosons}

\begin{figure}
\begin{center}
\includegraphics[width=0.28\textwidth]{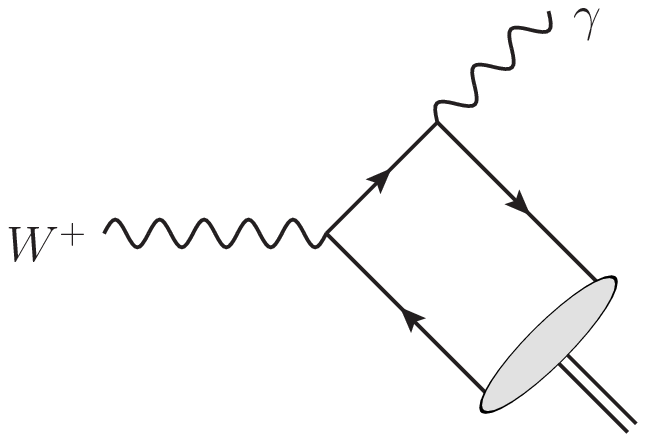}
\includegraphics[width=0.28\textwidth]{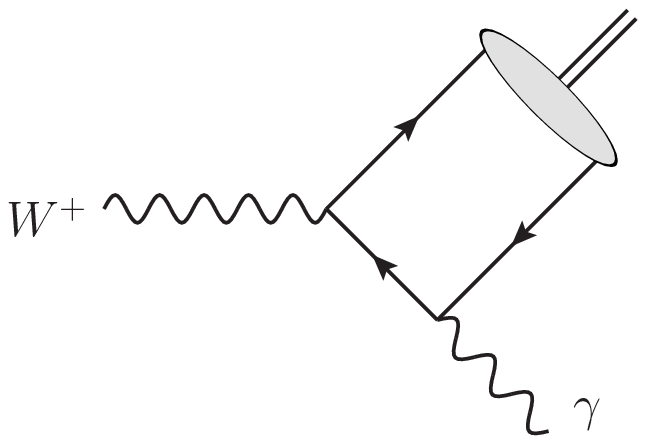}
\includegraphics[width=0.28\textwidth]{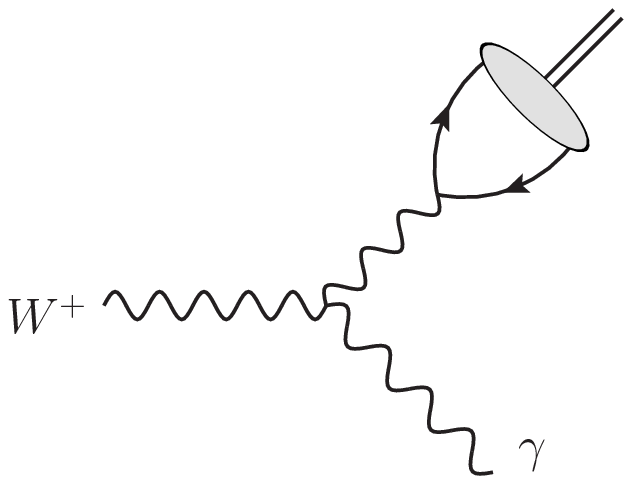}
\parbox{15.5cm}
{\caption{\label{fig:Wdiags}
Non-local (left and center) and local (right) contributions to the $W^+\to M^+\gamma$ decay amplitudes.}}
\end{center}
\end{figure}

The exclusive radiative decays $W^+\to M^+\gamma$ are, at first sight, very similar to the decays $Z^0\to M^0\gamma$. Indeed, the contributions from the first two diagrams shown in Figure~\ref{fig:Wdiags} can be obtained from the corresponding contributions to the $Z$-boson decay amplitudes by means of simple substitutions. The charged currents are flavor non-diagonal, and hence the final-state meson $M^+$ has a definite flavor structure described by a wave function $|u_i\bar d_j\rangle$. Note that now different electric-charge factors arise, depending on whether the photon is attached to the up-type quark or down-type anti-quark. The charged currents are purely left-handed, and hence we must replace 
\begin{equation}
   v_q,\,a_q \to \frac{\cos\theta_W}{2\sqrt2}\,V_{ij}
\end{equation}
in the equations of the previous section. However, a careful analysis shows that the first two diagrams in Figure~\ref{fig:Wdiags} give rise to an extra contribution with a different tensor structure. It reads
\begin{equation}\label{extra}
   i\Delta{\cal A}(W^+\to M^+\gamma)
   = \mp \frac{e g f_M}{2\sqrt2}\,V_{ij} \left( Q_u - Q_d \right)
    \frac{k\cdot\varepsilon_\gamma^*\,q\cdot\varepsilon_W}{k\cdot q} \,,
\end{equation}
where the upper (lower) sign refers to the case of a pseudoscalar (longitudinally polarized vector) meson in the final state. Note that this contribution is independent of the LCDA of the final-state meson. It vanishes for an on-shell (transverse) photon, but is not compatible with $U(1)_{\rm em}$ gauge invariance. 

Since the $W$ boson has a direct coupling to the photon, an extra contribution to the $W^+\to M^+\gamma$ decay amplitudes exists, which arises from the third diagram in Figure~\ref{fig:Wdiags}, in which the final-state meson is produced by the conversion of an off-shell $W$ boson. This graph has no analog in the $Z$-boson case. The corresponding contribution to the decay amplitude involves the meson matrix element of a local current, which to all orders in QCD is given in terms of a meson decay constant. We find
\begin{eqnarray}\label{Alocal}
   i{\cal A}_{\rm local}(W^+\to P^+\gamma)
   &=& \frac{eg f_P}{2\sqrt2}\,V_{ij}\,\varepsilon_W\cdot\varepsilon_\gamma^* \,, \\
   i{\cal A}_{\rm local}(W^+\to V^+\gamma)
   &=& - \frac{eg f_V}{2\sqrt2}\,V_{ij}\,\frac{2m_V}{m_W^2-m_V^2}\,\Big(
    q\cdot\varepsilon_V^*\,\varepsilon_W\cdot\varepsilon_\gamma^* 
    - k\cdot\varepsilon_\gamma^*\,\varepsilon_W\cdot\varepsilon_V^* 
    - q\cdot\varepsilon_W\,\varepsilon_\gamma^*\cdot\varepsilon_V^* \Big) \,, \nonumber
\end{eqnarray}
where we keep the exact dependence on the vector-meson mass $m_M$ for the time being. The second relation can be simplified by considering the cases of longitudinal and transverse polarization separately. The polarization vector for a longitudinally polarized vector meson can be decomposed as
\begin{equation}\label{epsdecomp}
   \varepsilon_V^{\parallel\mu}
   = \frac{1}{m_V} \left( k^\mu - \frac{2m_V^2}{m_W^2-m_V^2}\,q^\mu \right) ,
\end{equation}
which satisfies the conditions $k\cdot\varepsilon_V^\parallel=0$ and $(\varepsilon_V^\parallel)^2=-1$. The polarization vector for a transversely polarized vector meson is defined such that $k\cdot\varepsilon_V^\perp=q\cdot\varepsilon_V^\perp=0$. We then obtain
\begin{equation}\label{AlocalV}
\begin{aligned}
   i{\cal A}_{\rm local}(W^+\to V_\parallel^+\gamma)
   &= - \frac{eg f_V}{2\sqrt2}\,V_{ij} \left[ \varepsilon_W\cdot\varepsilon_\gamma^* 
    + {\cal O}\bigg( \frac{m_V^2}{m_W^2} \bigg) \right] , \\
   i{\cal A}_{\rm local}(W_\parallel^+\to V_\perp^+\gamma)
   &= - \frac{eg f_V}{2\sqrt2}\,V_{ij}\,\frac{m_V}{m_W}\,
    \varepsilon_\gamma^{\perp *}\cdot\varepsilon_V^{\perp *} \,. 
\end{aligned}
\end{equation}
The second amplitude is non-zero only if the $W$ boson is longitudinally polarized, and we have used a decomposition analogous to (\ref{epsdecomp}) to replace $-2q\cdot\varepsilon_W=\frac{m_W^2-m_V^2}{m_V}$ in the final result. The local amplitudes for $M=P,V_\parallel$ are such that they combine with the extra term in (\ref{extra}) to give a gauge-invariant result proportional to $\varepsilon_W^\perp\cdot\varepsilon_\gamma^{\perp *}$ \cite{Arnellos:1981gy,Manohar:1990hu}.

It follows from this discussion that, in analogy with (\ref{ampl1}), the leading-power amplitudes for the decays $W^+\to M^+\gamma$ can be written in the general form
\begin{equation}\label{ampl2}
   i{\cal A}(W^+\to M^+\gamma)
   = \pm\frac{eg f_M}{4\sqrt2}\,V_{ij} \left( i\epsilon_{\mu\nu\alpha\beta}\,
    \frac{k^\mu q^\nu\varepsilon_W^\alpha\,\varepsilon_\gamma^{*\beta}}{k\cdot q}\,F_1^M 
    - \varepsilon_W^\perp\cdot\varepsilon_\gamma^{\perp *}\,F_2^M \right) .
\end{equation}
Summing (averaging) over the polarization states of the photon ($W$ boson), we obtain the corresponding decay rates
\begin{equation}\label{Wrates}
   \Gamma(W^+\to M^+\gamma) = \frac{\alpha m_W f_M^2}{48v^2}\,|V_{ij}|^2
    \left( \left| F_1^M \right|^2 + \left| F_2^M \right|^2 \right) .
\end{equation}
In close analogy with (\ref{FVPres}), we find that the form factors are given by
\begin{eqnarray}\label{HVP}
   F_1^M &=& Q_u\,I_+^M(m_W) + Q_d\,\bar I_+^M(m_W) 
    = \sum_{n=0}^\infty \left[ C_{2n}^{(+)}(m_W,\mu)\,a_{2n}^M(\mu) 
     - 3 C_{2n+1}^{(+)}(m_W,\mu)\,a_{2n+1}^M(\mu) \right] , \nonumber \\ 
   F_2^M &=& - 2 \left( Q_u - Q_d \right) + Q_u\,I_-^M(m_W) - Q_d\,\bar I_-^M(m_W) \\
   &=& - 2 + \sum_{n=0}^\infty \left[ 3 C_{2n}^{(-)}(m_W,\mu)\,a_{2n}^M(\mu) 
     - C_{2n+1}^{(-)}(m_W,\mu)\,a_{2n+1}^M(\mu) \right] . \nonumber
\end{eqnarray}
The contribution $-2$ to $F_2^M$ arises from the local contribution in (\ref{Alocal}). Corrections to these results are suppressed by $(\Lambda_{\rm QCD}/m_W)^2$ or $(m_M/m_W)^2$. The corresponding amplitudes for the decays $W^-\to M^-\gamma$ are obtained by replacing $V_{ij}\to V_{ij}^*$ in (\ref{ampl2}) and by replacing the charge factors $Q_u\leftrightarrow Q_d$ in (\ref{HVP}). In addition, one must take into account that the odd Gegenbauer moments of the meson $M^-$ have the opposite sign as those of $M^+$. This can be accounted for by replacing $I_\pm^M\leftrightarrow\bar I_\pm^M$. As a result, the form factor $F_1^M$ remains invariant, while the form factor $F_2^M$ changes sign. The decay rate in (\ref{Wrates}) stays invariant under these replacements. At low values of the factoriszation scale $\mu$, the Wilson coefficients $C_n^{(\pm)}(m_W,\mu)$ in (\ref{HVP}) contain large logarithms of the form $\ln(m_W^2/\mu^2)$, which can be resummed to all orders in perturbation theory by evaluating the scale-invariant quantities in (\ref{HVP}) at the scale $\mu=m_W$. We obtain
\begin{equation}
\begin{aligned}
   \mbox{Re}\,F_1^M &= 0.94 - 2.98\,a_1^M(m_W) + 1.05\,a_2^M(m_W) - 3.31\,a_3^M(m_W) 
    + 1.15\,a_4^M(m_W) \mp \dots \\
   &= 0.94 - 1.65\,a_1^M(\mu_0) + 0.42\,a_2^M(\mu_0) - 1.03\,a_3^M(\mu_0) 
    + 0.30\,a_4^M(\mu_0) \mp \dots \,, \\
   \mbox{Re}\,F_2^M &= 0.85 - 1.00\,a_1^M(m_W) + 3.16\,a_2^M(m_W) - 1.11\,a_3^M(m_W)
    + 3.45\,a_4^M(m_W) \mp \dots \\
   &= 0.85 - 0.55\,a_1^M(\mu_0) + 1.25\,a_2^M(\mu_0) - 0.34\,a_3^M(\mu_0)
    + 0.89\,a_4^M(\mu_0) \mp \dots \,.
\end{aligned}
\end{equation}

Decays into a transversely polarized vector meson are once again only allowed at twist-3 order. We decompose the decay amplitude in a form analogous to (\ref{amplperp}), such that
\begin{equation}\label{amplperp2}
   i{\cal A}(W_\parallel^+\to V_\perp^+\gamma)
   = - \frac{eg f_V}{4\sqrt2}\,V_{ij}\,\frac{m_V}{m_W} \left( i\epsilon_{\mu\nu\alpha\beta}\,
    \frac{k^\mu q^\nu\varepsilon_V^{*\alpha} \varepsilon_\gamma^{*\beta}}{k\cdot q} F_1^\perp 
    - \varepsilon_V^{\perp *}\cdot\varepsilon_\gamma^{\perp *}\,F_2^\perp \right) .
\end{equation}
The corresponding decay rate, summed (averaged) over the polarizations of the final-state (initial-state) particles, is given by
\begin{equation}
   \Gamma(W\to V_\perp\gamma) = \frac{\alpha m_W f_V^2}{48v^2}\,|V_{ij}|^2\,\frac{m_V^2}{m_W^2}
    \left( \left| F_1^\perp \right|^2 + \left| F_2^\perp \right|^2 \right) .
\end{equation}
The form factors can be calculated in analogy with the discussion in the previous section. The final results are
\begin{equation}\label{Hperpres2}
\begin{aligned}
   F_1^\perp &= - 2 \int_0^1\!dx \left( \frac{Q_u\ln x}{1-x}
    + \frac{Q_d\ln(1-x)}{x} \right) \phi_V(x,\mu) \,, \\
   F_2^\perp &= - 4 \left( Q_u - Q_d \right) - 2 \int_0^1\!dx \left( \frac{Q_u\ln x}{1-x}
    - \frac{Q_d\ln(1-x)}{x} \right) \phi_V(x,\mu) \,,
\end{aligned}
\end{equation}
where for simplicity we neglect contributions proportional to the quark masses. The local contribution in (\ref{AlocalV}) adds $-2(Q_u-Q_d)$ to $F_2^\perp$. When expressed in terms of Gegenbauer moments, these results take the form
\begin{equation}
\begin{aligned}
   F_1^\perp &= 1 + \sum_{n=1}^\infty \frac{a_{2n}^V(\mu)}{(n+1)(2n+1)} 
    - \sum_{n=0}^\infty \frac{3 a_{2n+1}^V(\mu)}{(n+1)(2n+3)} \,, \\
   F_2^\perp &= -1 + \sum_{n=1}^\infty \frac{3 a_{2n}^V(\mu)}{(n+1)(2n+1)}  
    - \sum_{n=0}^\infty \frac{a_{2n+1}^V(\mu)}{(n+1)(2n+3)} \,.
\end{aligned}
\end{equation}
In the limit where the Gegenbauer moments are neglected, we find in analogy with (\ref{smallrat}) that the ratio 
\begin{equation}
   \frac{\Gamma(W^+\to V_\perp^+\gamma)}{\Gamma(W^+\to V_\parallel^+\gamma)} 
   \approx \frac{m_V^2}{m_W^2} \,.
\end{equation}
is strongly suppressed.

\subsection{Absence of enhanced contributions from the axial anomaly}
\label{sec:anomaly}

In has been suggested in the literature that the decay amplitudes for $Z\to P\gamma$ and $W\to P\gamma$ receive a very large enhancement due to an analog of the axial anomaly, which gives a contribution $\sqrt{2}\alpha/(\pi f_\pi)$ to the $\pi^0\to\gamma\gamma$ decay amplitude that does not vanish in the chiral limit \cite{Jacob:1989pw,Keum:1993eb}. We now explain why such a contribution does not exist in our case.

\begin{figure}
\begin{center}
\includegraphics[width=0.3\textwidth]{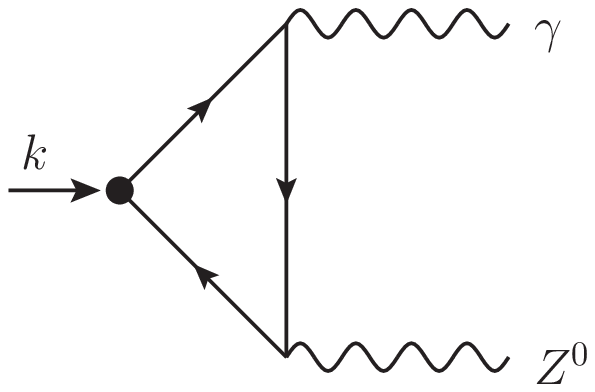}
\includegraphics[width=0.3\textwidth]{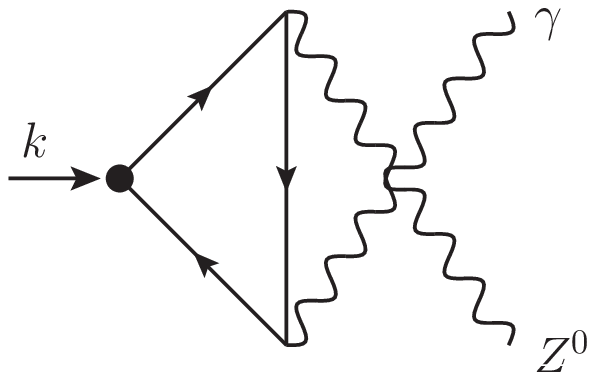}
\parbox{15.5cm}
{\caption{\label{fig:triangle}
One-loop triangle graphs giving rise to the axial anomaly. The dot represents the axial-vector current $A_q^\mu$.}}
\end{center}
\end{figure}

We consider the axial current $A_q^\mu=\bar q\gamma^\mu\gamma_5 q$ and evaluate the triangle diagrams shown in Figure~\ref{fig:triangle}, where instead of two photons we take the external particles to be a photon and a $Z$ boson. When taking the divergence of the current and considering the chiral limit $m_q\to 0$, we find that the loop diagrams vanish if one naively assumes that $\gamma_5$ anti-commutes with all Dirac matrices $\gamma^\mu$. However, a more careful regularization prescription adopting the HV scheme shows, like in the case of two external photons, that a finite remainder exists. It corresponds to a local operator for two gauge bosons. In operator language, we find that
\begin{equation}
   \partial_\mu A_q^\mu = 2im_q\,\bar q\gamma_5 q 
    - \frac{N_c\,\alpha}{4\pi}\,\epsilon_{\mu\nu\alpha\beta} 
    \left( Q_q^2\,F^{\mu\nu} F^{\alpha\beta} + \frac{2Q_q v_q}{\sin\theta_W\cos\theta_W}\,
    F^{\mu\nu} Z^{\alpha\beta} + \dots \right) ,
\end{equation}
where for completeness we have included the terms proportional to the quark mass on the right-hand side. We do not show contributions involving two electroweak gauge bosons ($ZZ$ or $W^+ W^-$) or two gluons, since they are irrelevant to our discussion. After electroweak symmetry breaking, the heavy $Z$ boson acts like an external source, which is invariant under $U(1)_{\rm em}$ gauge transformations. We see that anomalous contributions to the divergence of the axial-vector current not only involve the photon field, but also the $Z$ boson. Indeed, such anomalous terms even arise for the charged currents $A_{ij}^\mu=\bar d_j\gamma^\mu\gamma_5 u_i$. Evaluating the corresponding triangle graphs with an external $\gamma W$ state, we obtain
\begin{equation}
   \partial_\mu A_{ij}^\mu = i(m_{u_i}+m_{d_j})\,\bar d_j\gamma_5 u_i
    + ie\,\bar d_j\rlap{\hspace{1mm}/}{A}\gamma_5 u_i
    - \frac{N_c\,\alpha}{4\pi}\,\frac{V_{ij}}{3\sqrt{2}\sin\theta_W}\,
    \epsilon_{\mu\nu\alpha\beta}\,F^{\mu\nu} W^{+\alpha\beta} + \dots \,,
\end{equation}
where we omit a contribution involving two electroweak gauge bosons ($WZ$).

\begin{figure}
\begin{center}
\includegraphics[width=0.3\textwidth]{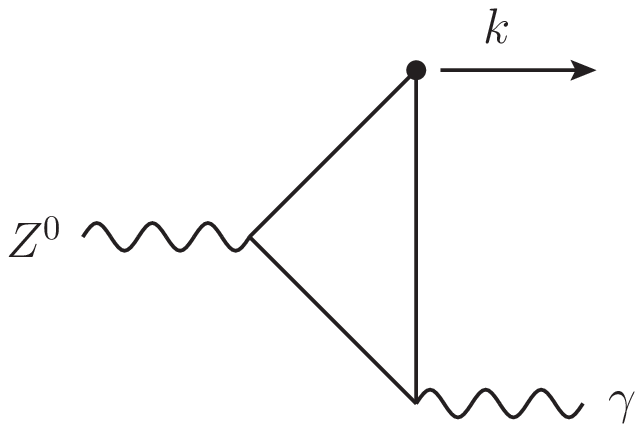}
\includegraphics[width=0.27\textwidth]{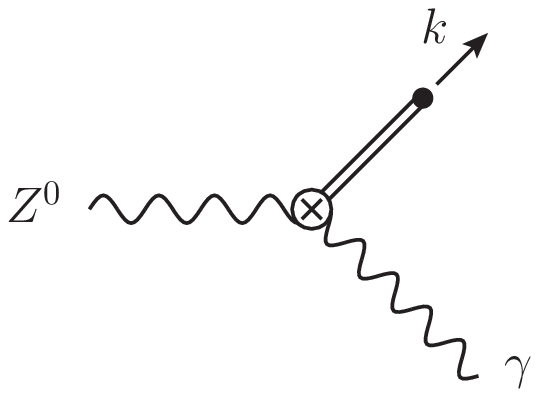}
\parbox{15.5cm}
{\caption{\label{fig:trianglecontr}
Left: One-loop diagram contributing to the amplitude ${\cal M}^{\mu\alpha\beta}$ describing the decay of a $Z$ boson into a photon and an axial current $A_q^\mu$. Right: A hypothetical anomaly-mediated contribution to the $Z^0\to P\gamma$ decay amplitude. The crossed circle represents the decay amplitude in (\ref{anomalyA}); the double line shows the meson propagator.}}
\end{center}
\end{figure}

Does the existence of the anomalous di-boson terms in the above relations imply that there exist  enhanced contributions to the $Z\to P\gamma$ and $W\to P\gamma$ decay amplitudes, in analogy to the famous case of the $\pi^0\to\gamma\gamma$ amplitude? This possibility was suggested in \cite{Jacob:1989pw,Keum:1993eb}, where the authors speculated about a huge enhancement of the rates for the radiative decays $Z\to\pi^0\gamma$ and $W^+\to D_s\gamma$ (see also the recent paper \cite{Mangano:2014xta}, which considers the decay $W^+\to\pi^-\gamma$). We now demonstrate that such an enhanced contribution does not exist, focussing for concreteness on the case of a neutral pseudoscalar meson meson $P$. Let us parameterize a hypothetical anomalous contribution to the $Z\to P\gamma$ decay amplitude in the form
\begin{equation}\label{anomalyA}
   i{\cal A}(Z\to P\gamma) = iA\,\epsilon_{\mu\nu\alpha\beta}\,k^\mu q^\nu
    \varepsilon_Z^\alpha\,\varepsilon_\gamma^{*\beta} \,,
\end{equation}
where the general structure of the amplitude is consistent with (\ref{ampl1}). Let us furthermore parameterize the amplitude coupling an initial-state $Z$ boson to the axial current $A_q^\mu$ and a photon as
\begin{equation}
   i{\cal M}^\mu(k,q)\equiv i{\cal M}^{\mu\alpha\beta}(k,q)\,
    \varepsilon_Z^\alpha\,\varepsilon_\gamma^{*\beta} \,.
\end{equation}
At lowest order, this amplitude is obtained from the diagram shown on the left in Figure~\ref{fig:trianglecontr}. Inserting a complete set of hadron states that can be interpolated by the axial current $A_q^\mu$, and summing over quark flavors, we find that the amplitude $i{\cal M}^{\mu\alpha\beta}(k,q)$ contains the following contribution from the the single-hadron state $P$:
\begin{equation}\label{anom_contrib}
   i{\cal M}^{\mu\alpha\beta}(k,q) \ni \sum_q\,c_q^P\,(-if_P k^\mu)\,
    \frac{i}{k^2-m_P^2}\,iA\,\epsilon_{\rho\sigma\alpha\beta}\,k^\rho q^\sigma
   \to i\sum_q\,c_q^P\,\frac{f_P A}{k^2}\,k^\mu\epsilon_{\rho\sigma\alpha\beta}\,
    k^\rho q^\sigma \,.
\end{equation}
This can be read off from the graph shown on the right in the figure. In the last step we have taken the chiral limit, in which the meson becomes massless. The key feature of the anomalous contribution would be that it exhibits a $1/k^2$ pole in this limit. We have calculated the one-loop contributions to the amplitude ${\cal M}^{\mu\alpha\beta}$ by evaluating the loop diagrams in Figure~\ref{fig:trianglecontr}. We find that the contribution associated with the tensor structure shown in (\ref{anom_contrib}) is proportional to the expression (with $p=k+q$)
\begin{equation}\label{curious}
   \frac{\Gamma(1+\epsilon)\,\Gamma^2(1-\epsilon)}{\Gamma(2-2\epsilon)}\,
    \frac{\mu^{2\epsilon}}{\left(p^2-k^2\right)^2} \left[ p^2\,
    \frac{\left(-p^2\right)^{-\epsilon} - \left(-k^2\right)^{-\epsilon}}{\epsilon}\,
    + \frac{\left(-k^2\right)^{1-\epsilon} - \left(-p^2\right)^{1-\epsilon}}{1-\epsilon} \right] ,
\end{equation}
where $p^2\equiv p^2+i0$ and $k^2\equiv k^2+i0$. For the case of the $\pi^0\to\gamma\gamma$ amplitude $p^2=0$ for the external photon, and in the limit $\epsilon\to 0$ one obtains $-1/k^2$, which indeed exhibits a pole. Form the residue of this pole one can derive the anomaly-mediated $\pi^0\to\gamma\gamma$ decay amplitude. For our case, on the other hand, $p^2=m_Z^2$ is equal to the mass of the decaying heavy gauge boson, in which case the above expression does not exhibit a $1/k^2$ pole, but is instead proportional to $1/m_Z^2$. Hence we conclude that $A=0$ in (\ref{anom_contrib}). Note that in the limit $k^2\to 0$ one obtains from (\ref{curious})  
\begin{equation}
   \frac{1}{m_Z^2} \left( \frac{1}{\epsilon} + \ln\frac{m_Z^2}{\mu^2} - i\pi
    + \mbox{const.} \right) ,
\end{equation}
which is precisely of the form of our (bare) hard-scattering coefficients.

\subsection{Phenomenological results}
\label{sec:numerics}

\begin{table}
\begin{center}
\begin{tabular}{|c|c||c|c|}
\hline 
Decay mode & Branching ratio & asymptotic\ & ~~LO~~ \\ 
\hline 
$Z^0\to\pi^0\gamma$ & $(9.80\,_{-\,0.14}^{+\,0.09}\,{}_\mu\pm 0.03_{f}
 \pm 0.61_{a_2}\pm 0.82_{a_4})\cdot 10^{-12}$ & 7.71 & 14.67 \\ 
$Z^0\to\rho^0\gamma$ & $(4.19\,_{-\,0.06}^{+\,0.04}\,{}_\mu\pm 0.16_{f}
 \pm 0.24_{a_2}\pm 0.37_{a_4})\cdot 10^{-9}$ & 3.63 & 5.68 \\ 
$Z^0\to\omega\gamma$ & $(2.82\,_{-\,0.04}^{+\,0.03}\,{}_\mu\pm 0.15_{f}
 \pm 0.28_{a_2}\pm 0.25_{a_4})\cdot 10^{-8}$ & 2.48 & 3.76 \\ 
$Z^0\to\phi\gamma$ & $(1.04\,_{-\,0.02}^{+\,0.01}\,{}_\mu\pm 0.05_{f}
 \pm 0.07_{a_2}\pm 0.09_{a_4})\cdot 10^{-8}$ & 0.86 & 1.49 \\
$Z^0\to J/\psi\,\gamma$ & $(8.02\,_{-\,0.15}^{+\,0.14}\,{}_\mu\pm 0.20_{f}
 \,_{-\,0.36}^{+\,0.39}\,{}_\sigma)\cdot 10^{-8}$ & 10.48 & 6.55 \\
$Z^0\to\Upsilon(1S)\,\gamma$ & $(5.39\,_{-\,0.10}^{+\,0.10}\,{}_\mu\pm 0.08_{f}
 \,_{-\,0.08}^{+\,0.11}\,{}_\sigma)\cdot 10^{-8}$ & 7.55 & 4.11 \\
$Z^0\to\Upsilon(4S)\,\gamma$ & $(1.22\,_{-\,0.02}^{+\,0.02}\,{}_\mu\pm 0.13_{f}
 \,_{-\,0.02}^{+\,0.02}\,{}_\sigma)\cdot 10^{-8}$ & 1.71 & 0.93 \\
$Z^0\to\Upsilon(nS)\,\gamma$ & $(9.96\,_{-\,0.19}^{+\,0.18}\,{}_\mu\pm 0.09_{f}
 \,_{-\,0.15}^{+\,0.20}\,{}_\sigma)\cdot 10^{-8}$ & 13.96 & 7.59 \\
\hline 
\end{tabular}
\parbox{15.5cm}
{\caption{\label{tab:BRsZ} 
Predicted branching fractions for various $Z\to M\gamma$ decays, including error estimates due to scale dependence (subscript ``$\mu$'') and the uncertainties in the meson decay constants (``$f$''), the Gegenbauer moments of light mesons (``$a_n$''), and the width parameters of heavy mesons (``$\sigma$''). See text for further explanations.}}
\end{center}
\end{table} 

We are now ready to present detailed numerical predictions for the various radiative decay modes. We start with the decays of the $Z$ boson, using relation (\ref{Zrates}). Besides the input parameters already mentioned, we need the $Z$-boson mass $m_Z=(91.1876\pm 0.0021)$\,GeV and total width $\Gamma_Z=(2.4955\pm 0.0009)$\,GeV \cite{Agashe:2014kda}. When squaring the decay amplitudes, we expand the resulting expressions consistently to first order in $\alpha_s$. The imaginary parts of the form factors in (\ref{FVPres}) do not enter at this order. Our results are presented in Table~\ref{tab:BRsZ}. Significant uncertainties in our predictions arise from the hadronic input parameters, in particular the meson decay constants (see Appendix~\ref{app:decay_constants}) and the various Gegenbauer moments. Their impact is explicitly shown in the table. Our error budget also includes a perturbative uncertainty, which we estimate by varying the factorization scale by a factor of~2 about the default value $\mu=m_Z$. All other uncertainties, such as those in the values of Standard Model parameters, are negligible. Note also that power corrections from higher-twist LCDAs are bound to be negligibly small, since they scale like $(\Lambda_{\rm QCD}/m_Z)^2$ for light mesons and at most like $(m_M/m_Z)^2$ for heavy ones. The predicted branching fractions range from about $10^{-11}$ for $Z^0\to\pi^0\gamma$ to about $10^{-7}$ for $Z^0\to J/\psi\,\gamma$. In the last row, the symbol $\Upsilon(nS)$ means that we sum over the first three $\Upsilon$ states ($n=1,2,3$). Strong, mode-specific differences arise foremost from the relevant flavor-dependent coefficients in Table~\ref{tab:coefs}, as well as from differences in the values of the decay constants. The combined uncertainties in the predictions for the branching fractions are typically of order 10\% and are dominated by the uncertainties in the shapes of the LCDAs. The only exception are the decays $Z^0\to\Upsilon\gamma$, for which the relevant hadronic overlap integral is constrained by the model-independent relation (\ref{eq16}).

\begin{figure}
\begin{center}
\includegraphics[width=0.34\textwidth]{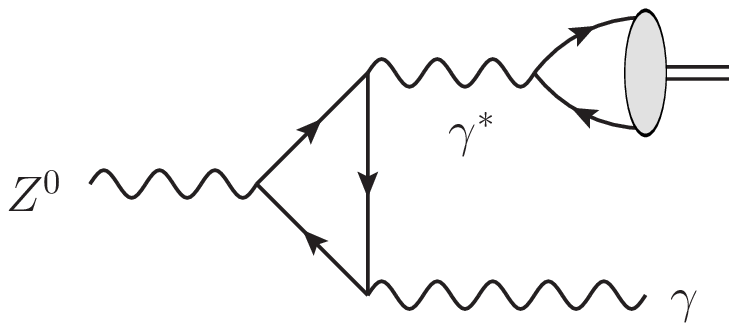}
\includegraphics[width=0.3\textwidth]{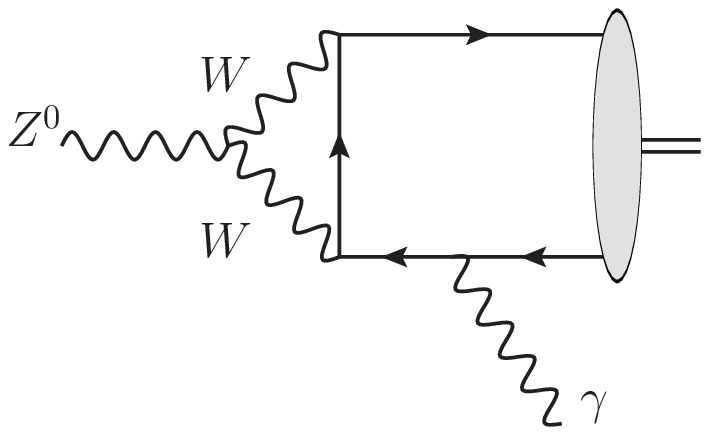}
\includegraphics[width=0.3\textwidth]{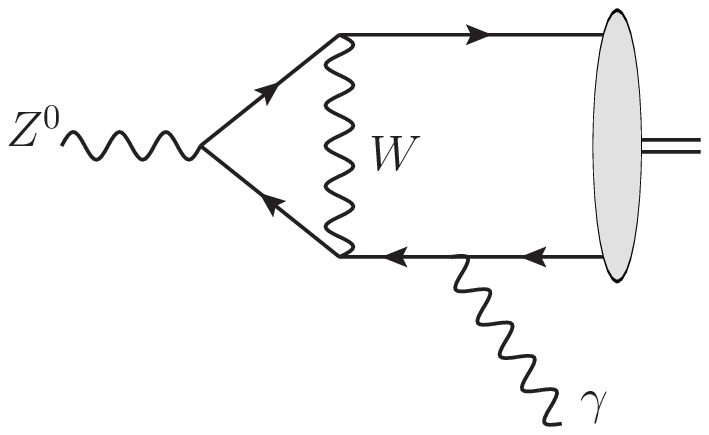}
\parbox{15.5cm}
{\caption{\label{fig:EWcors}
Examples of QED (left) and electroweak radiative corrections (center and right) to the $Z\to M\gamma$ decay amplitudes. The last two diagrams can give rise to flavor-violating decays in the Standard Model.}}
\end{center}
\end{figure}

In our analysis we neglect two-loop QCD corrections, whose effects should be covered by the error we estimate from scale variations, and one-loop QED or electroweak radiative corrections, a few examples of which are shown in Figure~\ref{fig:EWcors}. Their impact should be much smaller than the theoretical uncertainties inherent in our predictions. Consider, as a concrete example, the contribution of the first diagram, which only contributes to the $Z\to V\gamma$ amplitudes. Notice that the photon propagator $1/k^2$ with $k^2=m_V^2$ is cancelled, because the $Z\to\gamma\gamma^*$ amplitude vanishes if both photons are on-shell \cite{Hagiwara:1990dx}. As a result, there is no enhancement factor and the diagram is suppressed, compared with the leading contributions shown in Figure~\ref{fig:LOgraphs}, by a factor $\alpha/\pi\sim 2\cdot 10^{-3}$. This naive estimate is confirmed by the result of a detailed calculation of this contribution to the $Z\to J/\psi\,\gamma$ decay amplitude, which found that its effect leads to a reduction of the leading contribution by 0.2\%, corresponding to a 0.4\% correction of the branching ratio \cite{Huang:2014cxa}. In the same paper, the authors have presented predictions for three of the $Z\to V\gamma$ decay modes along with theoretical error estimates. They are $\mbox{Br}(Z\to\phi\gamma)=(11.7\pm 0.8)\cdot 10^{-9}$, $\mbox{Br}(Z\to J/\psi\,\gamma)=(9.96\pm 1.86)\cdot 10^{-8}$, and $\mbox{Br}(Z\to\Upsilon(1S)\,\gamma)=(4.93\pm 0.51)\cdot 10^{-8}$. The last two branching ratios are consistent with our findings within errors. Note that in the NRQCD approach adopted by these authors the decay constants of the heavy quarkonia are themselves derived from an expansion about the non-relativistic limit. This introduces additional uncertainties, which can be avoided if the decay constants are extracted from data, as discussed in Appendix~\ref{app:decay_constants}. The analysis of the decay $Z^0\to\phi\gamma$ presented in \cite{Huang:2014cxa} uses an approach similar to ours but only includes the leading logarithmic evolution effects from the hadronic scale $\mu_0=1$\,GeV to the high scale $\mu\sim m_Z$. Their result is consistent with ours but has a smaller uncertainty. The non-logarithmic ${\cal O}(\alpha_s)$ corrections included here for the first time reduce the branching ratio by a significant amount. We also find that the present ignorance about the precise shape of the $\phi$-meson LCDA gives rise to a larger uncertainty.

\begin{table}
\begin{center}
\begin{tabular}{|c|c||c|c|}
\hline 
Decay mode & Branching ratio & asymptotic\ & ~~LO~~ \\ 
\hline 
$W^\pm\to\pi^\pm\gamma$ & $(4.00\,_{-\,0.11}^{+\,0.06}\,{}_\mu\pm 0.01_{f}
 \pm 0.49_{a_2}\pm 0.66_{a_4})\cdot 10^{-9}$ & 2.45 & 8.09 \\ 
$W^\pm\to\rho^\pm\gamma$ & $(8.74\,_{-\,0.26}^{+\,0.17}\,{}_\mu\pm 0.33_{f}
 \pm 1.02_{a_2}\pm 1.57_{a_4})\cdot 10^{-9}$ & 6.48 & 15.12 \\ 
$W^\pm\to K^\pm\gamma$ & $(3.25\,_{-\,0.09}^{+\,0.05}\,{}_\mu\pm 0.03_{f}
 \pm 0.24_{a_1}\pm 0.38_{a_2}\pm 0.51_{a_4})\cdot 10^{-10}$ & 1.88 & 6.38 \\ 
$W^\pm\to K^{*\pm}\gamma$ & $(4.78\,_{-\,0.14}^{+\,0.09}\,{}_\mu\pm 0.28_{f}
 \pm 0.39_{a_1}\pm 0.66_{a_2}\pm 0.80_{a_4})\cdot 10^{-10}$ & 3.18 & 8.47 \\ 
$W^\pm\to D_s\gamma$ & $(3.66\,_{-\,0.07}^{+\,0.02}\,{}_\mu\pm 0.12_{\rm CKM}\pm 0.13_{f}
 \,_{-\,0.82}^{+\,1.47}\,{}_\sigma)\cdot 10^{-8}$ & 0.98 & 8.59 \\ 
$W^\pm\to D^\pm\gamma$ & $(1.38\,_{-\,0.02}^{+\,0.01}\,{}_\mu\pm 0.10_{\rm CKM}\pm 0.07_{f}
 \,_{-\,0.30}^{+\,0.50}\,{}_\sigma)\cdot 10^{-9}$ & 0.32 & 3.42 \\ 
$W^\pm\to B^\pm\gamma$ & $(1.55\,_{-\,0.03}^{+\,0.00}\,{}_\mu\pm 0.37_{\rm CKM}\pm 0.15_{f}
 \,_{-\,0.45}^{+\,0.68}\,{}_\sigma)\cdot 10^{-12}$ & 0.09 & 6.44 \\ 
\hline 
\end{tabular}
\parbox{15.5cm}
{\caption{\label{tab:BRsW} 
Predicted branching fractions for various $W\to M\gamma$ decays, including error estimates due to scale dependence and the uncertainties in the CKM matrix elements, the meson decay constants and the LCDAs. The notation is the same as in Table~\ref{tab:BRsZ}. See text for further explanations.}} 
\end{center}
\end{table} 

We now proceed to present our predictions for exclusive radiative decays of $W$ bosons. In this case we need the input parameters $m_W=(80.385\pm 0.015)$\,GeV and $\Gamma_W=(2.0897\pm 0.0008)$\,GeV, as well as the relevant entries of the quark mixing matrix, which are $|V_{ud}|=0.97425\pm 0.00022$, $|V_{us}|=0.2253\pm 0.0008$, $|V_{cs}|=0.986\pm 0.016$, $|V_{cd}|=0.225\pm 0.008$, $|V_{cb}|=(41.1\pm 1.3)\cdot 10^{-3}$, and $|V_{ub}|=(4.13\pm 0.49)\cdot 10^{-3}$ \cite{Agashe:2014kda}. Starting from relation (\ref{Wrates}), we obtain the results shown in Table~\ref{tab:BRsW}. In this case the pattern of the different decay modes reflects mainly the pattern of the relevant CKM matrix elements, and to a lesser extent the differences in the decay constants. The Cabibbo-allowed decays $W\to\pi\gamma, \rho\gamma$, and $D_s\gamma$ have branching fractions of order few times $10^{-9}$ to few times $10^{-8}$, where decays into heavy mesons are enhanced due to the structure of the relevant overlap integral in (\ref{eq18}). The Cabibbo-suppressed modes $W\to K^{(*)}\gamma$ and the strongly CKM-suppressed decay $W\to B\gamma$ have correspondingly smaller branching ratios. The uncertainties inherited from CKM elements are shown where they are significant. In a recent paper, the $W^\pm\to\pi^\pm\gamma$ branching ratio was estimated to be $0.64\cdot 10^{-9}$ \cite{Mangano:2014xta}, which is about 6.3 times smaller than the value we obtain (see below).

In the last two columns in Tables~\ref{tab:BRsZ} and \ref{tab:BRsW} we show different approximations to our results. The first one (labelled ``asymptotic'') gives the central values of the branching ratios (in the appropriate units) obtained if the asymptotic form $6x(1-x)$ of the meson LCDA is employed. As we have explained, RG evolution effects from the low hadronic scale $\mu_0=1$\,GeV up to the electroweak scale have the effect of strongly suppressing the contributions from higher Gegenbauer moments. Indeed, we observe that using the asymptotic form provides reasonable approximations in most cases (especially for the $Z\to M\gamma$ modes). The corresponding expressions for the decay rates read
\begin{equation}\label{Gamasymp}
\begin{aligned}
   \Gamma(Z^0\to M^0\gamma) \big|_{\rm asymp} 
   &= \frac{\alpha m_Z f_M^2}{6v^2}\,{\cal Q}_M^2
    \left[ 1 - \frac{10}{3}\,\frac{\alpha_s(m_Z)}{\pi} \right] , \\
   \Gamma(W^\pm\to M^\pm\gamma) \big|_{\rm asymp} 
   &= \frac{\alpha m_W f_M^2}{24v^2}\,|V_{ij}|^2
    \left[ 1 - \frac{17}{3}\,\frac{\alpha_s(m_W)}{\pi} \right] ,
\end{aligned}   
\end{equation}
where $V_{ij}$ is the relevant CKM matrix element for the production of the charged meson $M^+$. The dominant corrections to the $Z\to M\gamma$ branching fractions arise from the second Gegenbauer moment $a_2^M$, which is positive for light mesons and negative for heavy quarkonia. The dominant corrections to the $W\to M\gamma$ branching fractions with kaons, $D$ mesons or $B$ mesons in the final state arise from the first Gegenbauer moment $a_1^M$. It gives a large positive contribution in all cases. In the case of heavy mesons this effect is particularly pronounced. The approximate results in (\ref{Gamasymp}) are fully consistent with corresponding (tree-level) expressions derived in \cite{Arnellos:1981gy}. The result for the $Z^0\to\pi^0\gamma$ decay rate derived in \cite{Manohar:1990hu} is lower than ours by a factor 4/9, and the formula for the $W^\pm\to\pi^\pm\gamma$ decay rate derived in \cite{Mangano:2014xta} differs from (\ref{Gamasymp}) by a factor 2/9. The origin of the discrepancy is related to the fact that the theoretical approach used in these papers is based on an expansion in a parameter $\omega_0=2$, and the numerical estimates are obtained by keeping only the leading term in the expansion -- a fact that was admitted in these papers. In Appendix~\ref{app:manohar} we trace the source of the discrepancy in more detail. In the last column in Tables~\ref{tab:BRsZ} and \ref{tab:BRsW} (labelled ``LO'') we present the branching ratios one would obtain at tree level using the model predictions for the LCDAs at the low scale $\mu_0$. In this approximation the one-loop QCD corrections, which contain large logarithms of the form $\alpha_s\ln(m_{Z,W}^2/\mu_0^2)$, are omitted. For most decays the corresponding results overshoot the values obtained at next-to-leading order by significant amounts; only for decays into heavy quarkonia they underestimate the branching fractions. 

\begin{table}
\begin{center}
\begin{tabular}{|c|c|c|}
\hline 
Decay mode & Branching ratio & SM background \\ 
\hline
$Z^0\to K^0\gamma$ & $\big[ (7.70\pm 0.83)\,|v_{sd}|^2
 + (0.01\pm 0.01)\,|a_{sd}|^2 \big]\cdot 10^{-8} \phantom{\Big|}\! $
 & $\frac{\lambda}{\sin^2\theta_W}\,\frac{\alpha}{\pi}\sim 2\cdot 10^{-3}$ \\[1mm] 
$Z^0\to D^0\gamma$ & $\left[ (5.30\,_{-\,0.43}^{+\,0.67})\,|v_{cu}|^2
 + (0.62\,_{-\,0.23}^{+\,0.36})\,|a_{cu}|^2 \right] \cdot 10^{-7}$ 
 & $\frac{\lambda}{\sin^2\theta_W}\,\frac{\alpha}{\pi}\sim 2\cdot 10^{-3}$ \\[1mm] 
$Z^0\to B^0\gamma$ & $\left[ (2.08\,_{-\,0.41}^{+\,0.59})\,|v_{bd}|^2
 + (0.77\,_{-\,0.26}^{+\,0.38})\,|a_{bd}|^2 \right] \cdot 10^{-7}$ 
 & $\frac{\lambda^3}{\sin^2\theta_W}\,\frac{\alpha}{\pi}\sim 8\cdot 10^{-5}$ \\[1mm]  
$Z^0\to B_s\gamma$ & $\left[ (2.64\,_{-\,0.52}^{+\,0.82})\,|v_{bs}|^2
 + (0.87\,_{-\,0.33}^{+\,0.51})\,|a_{bs}|^2 \right] \cdot 10^{-7}$ 
 & $\frac{\lambda^2}{\sin^2\theta_W}\,\frac{\alpha}{\pi}\sim 4\cdot 10^{-4}$ \\[1mm] 
\hline 
\end{tabular}
\parbox{15.5cm}
{\caption{\label{tab:ZdecayNP} 
Branching fractions for FCNC transitions $Z\to M\gamma$, which could arise from physics beyond the Standard Model. The different theoretical uncertainties have been added in quadrature. The last column shows our estimates for the irreducible Standard Model background up to which one can probe the flavor-changing couplings $v_{ij}$ and $a_{ij}$. Here $\lambda\approx 0.2$ is the Wolfenstein parameter.}}
\end{center}
\end{table} 

Future precision measurements of the exclusive radiative decays $Z\to M\gamma$ would serve as powerful tests of the Standard Model and of the framework of QCD factorization. The branching ratios for decays into vector mesons shown in Table~\ref{tab:BRsZ} are proportional to $|a_q|^2$, where $a_q$ denote the axial-vector couplings of the $Z$ boson to the various quarks. The couplings $|a_b|$ and $|a_c|$ have been measured at LEP with 1\% accuracy \cite{ALEPH:2005ab}, but no similarly accurate direct measurements of the couplings to light quarks are available. Given the theoretical precision of our predictions, it would be possible to determine these couplings with about 6\% accuracy. The decays $Z\to M\gamma$ can in principle also be used to search for non-standard FCNC couplings of the $Z$ boson. If such couplings exist, then the diagrams shown in Figure~\ref{fig:LOgraphs} can lead to final-state mesons of mixed flavor, such as $K^0$, $D^0$, $B^0$ and $B_s$. It is straightforward to calculate the corresponding decay rates in our approach, starting from the general relations (\ref{Zrates}) and (\ref{FVPres}). We parameterize the non-standard vector and axial-vector couplings of the $Z$ boson by $v_{ij}$ and $a_{ij}$, respectively, where $i,j$ are the quark flavors of the final-state meson. Our predictions for the corresponding branching fractions are given in Table~\ref{tab:ZdecayNP}. At higher order some Standard Model background to these searches exists, since electroweak loop graphs such as those shown in the last two diagrams in Figure~\ref{fig:EWcors} can give rise to flavor-changing transitions. Naive dimensional analysis shows that the contributions of these diagrams, relative to the contributions from the graphs in Figure~\ref{fig:diags} (in units of the new-physics couplings $v_{ij}$ and $a_{ij}$), scale like $(\alpha/\pi)\,|V_{ik} V_{kj}^*|/\sin^2\theta_W$, where $k$ can be any one of the three possible generation indices. The relevant loop functions depend on the dimensionless ratios $m_k^2/m_Z^2$ and $m_W^2/m_Z^2$, which are either of ${\cal O}(1)$ or can be set to zero. Consequently one can only probe the new-physics couplings $v_{ij}$ and $a_{ij}$ up to some irreducible Standard Model background, which is estimated in the last column in Table~\ref{tab:ZdecayNP}. 

\begin{table}
\begin{center}
\begin{tabular}{|c|c||c|c|}
\hline 
$\big|\mbox{Re}\big[(v_{sd}\pm a_{sd})^2\big]\big|$ & $<2.9\cdot 10^{-8}$
 & $\big|\mbox{Re}\big[(v_{sd})^2-(a_{sd})^2\big]\big|$ & $<3.0\cdot 10^{-10}$ \\[1mm]
$\big|\mbox{Im}\big[(v_{sd}\pm a_{sd})^2\big]\big|$ & $<1.0\cdot 10^{-10}$ 
 & $\big|\mbox{Im}\big[(v_{sd})^2-(a_{sd})^2\big]\big|$ & $<4.3\cdot 10^{-13}$ \\[1mm] 
$\big|(v_{cu}\pm a_{cu})^2\big|$ & $<2.2\cdot 10^{-8}$
 & $\big|(v_{cu})^2-(a_{cu})^2\big|$ & $<1.5\cdot 10^{-8}$ \\[1mm] 
$\big|(v_{bd}\pm a_{bd})^2\big|$ & $<4.3\cdot 10^{-8}$
 & $\big|(v_{bd})^2-(a_{bd})^2\big|$ & $<8.2\cdot 10^{-9}$ \\[1mm]  
$\big|(v_{bs}\pm a_{bs})^2\big|$ & $<5.5\cdot 10^{-7}$
 & $\big|(v_{bs})^2-(a_{bs})^2\big|$ & $<1.4\cdot 10^{-7}$ \\ 
\hline
\end{tabular}
\parbox{15.5cm}
{\caption{\label{tab:UTbounds} 
Indirect constraints on the flavor-changing $Z$-boson couplings $v_{ij}$ and $a_{ij}$ (at 95\% confidence level) derived from neutral-meson mixing \cite{Bona:2007vi,Bertone:2012cu,Carrasco:2013zta}.}}
\vspace{-3mm}
\end{center}
\end{table} 

Possible FCNC couplings of the $Z$ boson are heavily constrained by precision flavor physics, in particular by bounds on the $\Delta F=2$ mixing amplitudes. It is a straightforward exercise to match our parameters $v_{ij}$ and $a_{ij}$ onto the Wilson coefficients in the general effective $\Delta F=2$ Hamiltonian as defined, e.g., in \cite{Bona:2007vi}. We obtain
\begin{equation}
   C_1 = \frac{4G_F}{\sqrt2} \left( v_{ij} + a_{ij} \right)^2 , \qquad
   \tilde C_1 = \frac{4G_F}{\sqrt2} \left( v_{ij} - a_{ij} \right)^2 , \qquad
   C_5 = - \frac{4G_F}{\sqrt2} \left( v_{ij}^2 - a_{ij}^2 \right) .
\end{equation}
All other coefficients are zero at tree level. Using the bounds compiled in \cite{Bona:2007vi} as well as updated results reported in \cite{Bertone:2012cu,Carrasco:2013zta}, we find the upper bounds on various combinations of $v_{ij}$ and $a_{ij}$ parameters shown in Table~\ref{tab:UTbounds}. The strongest bounds exist for the coefficients $C_5$ of mixed-chirality operators (right column). They can be avoided by assuming that $v_{ij}=\pm a_{ij}$, such that the flavor-changing couplings are either purely left-handed or purely right-handed. Under this assumption one finds from the table that $|v_{sd}|<8.5\cdot 10^{-5}$, $|v_{cu}|<7.4\cdot 10^{-5}$, $|v_{bd}|<1.0\cdot 10^{-4}$ and $|v_{bs}|<3.7\cdot 10^{-4}$, and the same bounds apply to $|a_{ij}|$. If these indirect bounds are used, then the branching fraction shown in Table~\ref{tab:ZdecayNP} are predicted to be at most a few times $10^{-15}$ (a few times $10^{-14}$ for the case of $Z^0\to B_s\gamma$), meaning that they will be unobservable at the LHC and all currently discussed future facilities. We find it nevertheless worthwhile to illustrate the general idea of such new-physics searches. First of all, it should be emphasized that the indirect bounds derived from $K\!-\!\bar K$, $D\!-\!\bar D$ and $B_{d,s}\!-\!\bar B_{d,s}$ mixing are to some extent model dependent, since one cannot tell whether the flavor violation originates from the couplings of the $Z$ boson or from some other new particle. It is conceivable that in some (admittedly fine-tuned) models flavor-violating couplings of the $Z$ boson can be compensated by the effects of some other, heavy boson. Also, in deriving the bounds on a particular Wilson coefficient $C_i$ one assumes that a single new-physics operator is present at a time and sets the coefficients of all other operators to zero. The method presented here, on the other hand, is unique in that it allows one (in principle) to probe for flavor-changing couplings of the $Z$ boson directly and in a model-independent way, based on tree-level couplings of an on-shell particle. It should thus be seen as a complementary way to search for such effects. This method can also be generalized to the interesting case of flavor-changing exclusive Higgs-boson decays \cite{Kagan:2014ila}, for which the corresponding indirect bounds have been studied in \cite{Harnik:2012pb}.

\section{\boldmath Weak radiative hadronic decays $Z^0\to M^+W^-$}
\label{sec:weakrad}

\begin{figure}
\begin{center}
\includegraphics[width=0.28\textwidth]{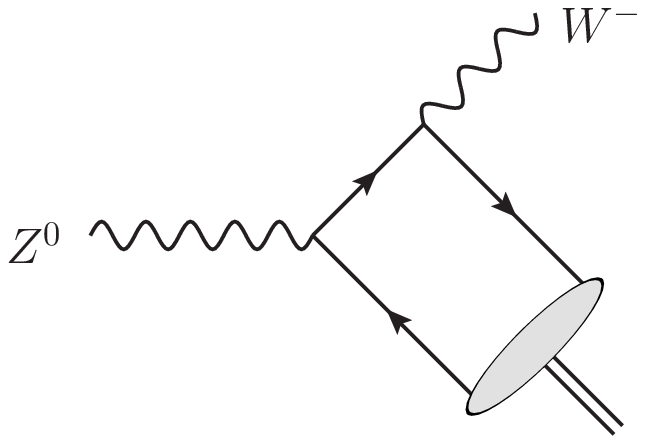}
\includegraphics[width=0.28\textwidth]{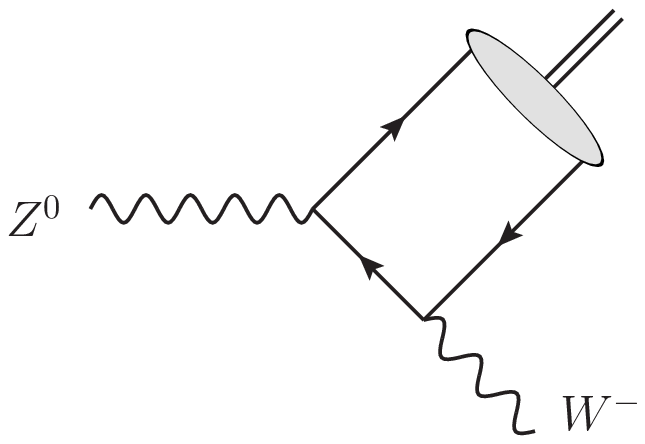}
\includegraphics[width=0.28\textwidth]{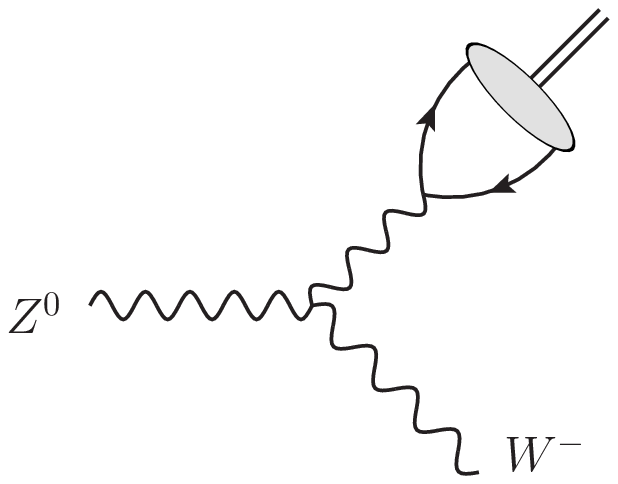}
\parbox{15.5cm}
{\caption{\label{fig:diags}
Non-local (left and center) and local (right) contributions to the $Z\to M^+W^-$ decay amplitudes.}}
\end{center}
\end{figure}

Exclusive decays of a $Z$ boson into a $W$ boson and a single meson $M$ are kinematically allowed as long as the final-state meson is lighter than the mass difference $m_Z-m_W\simeq 10.8$\,GeV. While similar at first sight to the radiative $Z$-boson decays studied in Section~\ref{sec:raddecays}, these decays are nevertheless interesting for several reasons. Unlike the photon, the final-state $W$ boson can be longitudinally polarized, and hence several different helicity amplitudes contribute to the decay. Also, the trilinear $ZWW$ coupling in the Standard Model gives rise to an additional contribution to the decay amplitude, in which the final-state meson is produced via the conversion of a $W$ boson. This term is analogous to the ``local'' contribution we encountered in our study of the radiative decays of $W$ bosons. Indeed, the leading-order Feynman graphs contributing to the $Z^0\to M^+W^-$ decay amplitudes, shown in Figure~\ref{fig:diags}, are analogous to those in Figure~\ref{fig:Wdiags}. Finally, and most interestingly, the decays $Z\to MW$ offer an opportunity to test the QCD factorization approach at a scale significantly lower than the $Z$-boson mass. A factorization theorem of the form (\ref{factorization}) can only be derived if the momentum of the final-state meson in the rest frame of the decaying particle is much larger than its mass, since only then the constituents of this meson can be described in terms of collinear quark and gluon fields in SCET. The relevant condition is 
\begin{equation}
   \frac{\lambda(m_Z^2,m_W^2,m_M^2)}{2m_Z} \gg m_M \,, 
    \quad \mbox{where} \quad
   \lambda(x,y,z)=\sqrt{(x-y-z)^2-4yz} \,.
\end{equation}
This condition is satisfied as long as $m_M\ll\frac{m_Z^2-m_W^2}{2m_Z}\approx 10.2$\,GeV. We can thus use the factorization approach to calculate the branching fractions for $Z\to MW$ decays with a light, strange or charm meson in the final state, but not with a $B$ meson. It also follows that the non-perturbative corrections to the factorization formula are organized in powers of $\big(\frac{\Lambda_{\rm QCD}\,m_Z}{m_Z^2-m_W^2}\big)^2$ rather than $(\Lambda_{\rm QCD}/m_Z)^2$. 

The local contribution to the decay amplitudes involves the meson matrix elements of local currents. In close analogy with (\ref{Alocal}), we find
\begin{eqnarray}\label{indirect}
   i{\cal A}_{\rm local}(Z\to P^+ W^-)
   &=& - \frac{g^2\cos\theta_W f_P}{2\sqrt2}\,V_{ij}\,\frac{m_Z^2-m_W^2}{m_W^2}\,
    \varepsilon_Z\cdot\varepsilon_W^* \,, \nonumber\\
   i{\cal A}_{\rm local}(Z\to V^+ W^-)
   &=& \frac{g^2\cos\theta_W f_V}{2\sqrt2}\,V_{ij}\,\frac{2m_V}{m_W^2-m_V^2} \\
   &&\mbox{}\times \Big( q\cdot\varepsilon_V^*\,\varepsilon_Z\cdot\varepsilon_W^* 
    - k\cdot\varepsilon_W^*\,\varepsilon_Z\cdot\varepsilon_V^* 
    + k\cdot\varepsilon_Z\,\varepsilon_W^*\cdot\varepsilon_V^* \Big) \,. \nonumber
\end{eqnarray}
These results are exact even as far as the dependence on the vector-meson mass is concerned. The second relation can be simplified by considering the cases of longitudinal and transverse polarization of the vector meson separately. The longitudinal polarization vector can be written as
\begin{equation}
   \varepsilon_V^{\parallel\mu}
   = \frac{1}{m_V}\,\frac{m_Z^2-m_W^2-m_V^2}{\lambda(m_Z^2,m_W^2,m_V^2)}
    \left( k^\mu - \frac{2m_V^2}{m_Z^2-m_W^2-m_V^2}\,q^\mu \right) .
\end{equation}
Using this result, the second amplitude can be simplified to read 
\begin{equation}\label{eq42}
   i{\cal A}(Z\to V_\parallel^+ W^-)
   = \frac{g^2\cos\theta_W f_V}{2\sqrt2}\,V_{ij}\,\frac{m_Z^2-m_W^2}{m_W^2}\,
    \varepsilon_Z\cdot\varepsilon_W^* \left[ 1 
    + {\cal O}\bigg( \frac{m_V^2 m_Z^2}{\left( m_Z^2-m_W^2 \right)^2} \bigg) \right] , 
\end{equation}
while the decay amplitudes for the processes $Z_\parallel\to V_\perp^+ W_\perp^-$ and $Z_\perp\to V_\perp^+ W_\parallel^-$, in which the final-state meson is transversely polarized, are suppressed relative to (\ref{eq42}) by factors of $m_V/m_Z$ and $m_V/m_W$, respectively. These power-suppressed amplitudes contribute to the decay rate at ${\cal O}(m_V^2/m_W^2)$ and are thus negligible. They will be neglected below. Up to very small corrections, the two amplitudes in (\ref{indirect}) thus have an identical structure. 

Similar to (\ref{ampl2}), we can write the most general form of the decay amplitudes in the form
\begin{equation}\label{ampl3}
\begin{aligned}
   i{\cal A}(Z\to M^+W^-)
   &= \pm\frac{g^2 f_M}{4\sqrt2\cos\theta_W}\,V_{ij} \left( 1 - \frac{m_W^2}{m_Z^2} \right) \\
   &\quad\times \left( i\epsilon_{\mu\nu\alpha\beta}\,
    \frac{k^\mu q^\nu\varepsilon_Z^\alpha\,\varepsilon_W^{*\beta}}{k\cdot q}\,F_1^M 
    - \varepsilon_Z\cdot\varepsilon_W^*\,F_2^M 
    + \frac{q\cdot\varepsilon_Z\,k\cdot\varepsilon_W^*}{k\cdot q}\,F_3^M \right) .   
\end{aligned}
\end{equation}
The last structure inside the parenthesis contributes only if the $W$ and $Z$ bosons are longitudinally polarized. It has no analog in the case of radiative $Z$ or $W$ decays.

Summing (averaging) over the polarizations of the final-state (initial-state) particles, and setting the meson mass to zero, we obtain from (\ref{ampl3}) the decay rate 
\begin{equation}\label{ZWMrate}
\begin{aligned}
   \Gamma(Z\to M^+W^-) 
   &= \frac{\alpha(m_Z)\,m_Z f_M^2}{48v^2}\,|V_{ij}|^2
    \left( 1 - \frac{m_W^2}{m_Z^2} \right)^2 \\
   &\mbox{}\times \left[ \left| F_1^M \right|^2 + \left| F_2^M \right|^2 
    + \frac{m_Z^2}{2m_W^2} 
    \left| \frac{m_Z^2+m_W^2}{2m_Z^2}\,F_2^M - \frac{m_Z^2-m_W^2}{2m_Z^2}\,F_3^M \right|^2 \right] .
\end{aligned}
\end{equation}
Notice the close similarity of this result with (\ref{Wrates}). The differences are that we must now evaluate the coupling $\alpha$ at the electroweak scale, where for simplicity we do not differentiate between $m_Z$ and $m_W$. Also, a phase-space suppression factor arises, and the third form factor $F_3^M$ yields a contribution that is absent in (\ref{Wrates}). In deriving the above relation we have used that $g^2=e^2/\sin^2\theta_W$ and $\sin^2\theta_W=1-m_W^2/m_Z^2$. 

Evaluating the first two Feynman graphs in Figure~\ref{fig:diags}, we obtain
\begin{equation}\label{ZWMffs}
\begin{aligned}
   F_1^M &= \int_0^1\!dx\,\phi_M(x,\mu) 
    \left[ \frac{Z_d}{x+(1-x)\,r} + \frac{Z_u}{(1-x)+xr} \right] , \\
   F_2^M &= 2 + \int_0^1\!dx\,\phi_M(x,\mu) 
    \left[ \frac{Z_d}{x+(1-x)\,r} - \frac{Z_u}{(1-x)+xr} \right] , \\
   F_3^M &= \int_0^1\!dx\,\phi_M(x,\mu)\,(1-2x)
    \left[ \frac{Z_d}{x+(1-x)\,r} + \frac{Z_u}{(1-x)+xr} \right] .
\end{aligned}
\end{equation}
Here $r=m_W^2/m_Z^2=\cos^2\theta_W$, and $Z_q=v_q+a_q$ are the left-handed couplings of quarks to the $Z$ boson. Explicitly, we have $Z_u=\frac12-\frac23\sin^2\theta_W$ and $Z_d=-\frac12+\frac13\sin^2\theta_W$. From the results shown in (\ref{indirect}) and (\ref{eq42}), we see that the local contribution from the third diagram in Figure~\ref{fig:diags} adds the term 2 to the form factors $F_2^M$. Note that the relevant combination
\begin{equation}
   \frac{1+r}{2}\,F_2^M - \frac{1-r}{2}\,F_3^M = 1
\end{equation}
entering the total decay rate is independent of the form of the LCDA. It follows that the third (longitudinal) term in the expression (\ref{ZWMrate}) for the decay rate yields $1/(2r)$. The integrals over the LCDA in our expressions for the form factors $F_{1,2}^M$ can be evaluated analytically to any given order in the Gegenbauer expansion, but in practice it is easier to evaluate them numerically. Because the functions in the denominators are slowly varying with $x$, we find that higher-order Gegenbauer moments in the expansion (\ref{Gegenbauer}) of the LCDA $\phi_M(x,\mu)$ give very small contributions. This is the essence of the approach by Manohar \cite{Manohar:1990hu}, which we illustrate in Appendix~\ref{app:manohar}. We obtain
\begin{equation}\label{H123}
\begin{aligned}
   \left| F_1^M \right|^2 + \left| F_2^M \right|^2 + \frac{1}{2r} 
   = 1.911 + 0.003\,a_1^M(\mu) - 0.011 a_2^M(\mu) + {\cal O}(10^{-3}) \,,
\end{aligned}
\end{equation}
where the leading term 1.911 can be written in closed form as
\begin{equation}\label{eq82}
   \left(1-2r+10r^2 \right)f^2(r) - 12r f(r) + 4 + \frac{1}{2r} \,, 
    \quad \mbox{with} \quad
   f(r) = \frac{1-r^2+2r\ln r}{(1-r)^3} \,.
\end{equation}
The calculation of ${\cal O}(\alpha_s)$ corrections to the results in (\ref{ZWMffs}) is an interesting project, which we leave for future work. We note, however, that in this case the value of the factorization scale should be taken lower than the mass of the decaying $Z$ boson. Two natural choices would be the typical momentum transfer, $\mu^2\sim 2k\cdot q=(m_Z^2-m_W^2)\approx (43\,\mbox{GeV})^2$, and (twice) the energy of the final-state meson in the $Z$-boson rest frame, $\mu\sim(m_Z^2-m_W^2)/m_Z\approx 20$\,GeV. In our analysis we shall use latter value, but since in the tree-level calculation presented here the scale dependence enters only through the tiny corrections involving $a_{1,2}^M(\mu)$ in (\ref{H123}) this choice has no noticeable effect.

\begin{table}
\begin{center}
\begin{tabular}{|c|c|}
\hline 
Decay mode & Branching ratio \\ 
\hline 
$Z^0\to\pi^\pm W^\mp$ & $(1.51\pm 0.005_{f})\cdot 10^{-10}$ \\ 
$Z^0\to\rho^\pm W^\mp$ & $(4.00\pm 0.15_{f})\cdot 10^{-10}$ \\ 
$Z^0\to K^\pm W^\mp$ & $(1.16\pm 0.01_{f})\cdot 10^{-11}$ \\ 
$Z^0\to K^{*\pm} W^\mp$ & $(1.96\pm 0.12_{f})\cdot 10^{-11}$ \\ 
$Z^0\to D_s W^\mp$ & $(6.04\pm 0.20_{\rm CKM}\pm 0.22_{f})\cdot 10^{-10}$ \\ 
$Z^0\to D^\pm W^\mp$ & $(1.99\pm 0.14_{\rm CKM}\pm 0.10_{f})\cdot 10^{-11}$ \\ 
\hline 
\end{tabular}
\parbox{15.5cm}
{\caption{\label{tab:BRsZWM} 
Predicted branching fractions for various $Z\to MW$ decays, including error estimates due to the uncertainties in the CKM matrix elements and the meson decay constants. Uncertainties in the shapes of the LCDAs have a negligible impact at tree level. Not shown are perturbative uncertainties due to the neglect of ${\cal O}(\alpha_s)$ corrections.}} 
\vspace{-6mm}
\end{center}
\end{table} 

Our predictions for the various $Z\to MW$ branching ratios are collected in Table~\ref{tab:BRsZWM}. They are smaller than the corresponding branching fractions for $W\to M\gamma$ decays by more than a factor 20 for light mesons and by more than a factor 60 for heavy mesons. Note, however, the curious fact that the $Z^0\to\pi^\pm W^\mp$ branching ratio is about 15 times larger than the $Z^0\to\pi^0\gamma$ branching ratio. The latter quantity is tiny, because it is proportional to ${\cal Q}_\pi^2\approx 7\cdot 10^{-4}$. We do not quote uncertainties related to the shapes of the meson LCDAs, which have a negligible impact on our result, see (\ref{H123}). We stress that to the quoted errors one should add an uncertainty accounting for our neglect of QCD radiative corrections. We expect that they can change the branching ratios by 10--20\%.

It is interesting to compare our results with those obtained by Manohar in \cite{Manohar:1990hu}. To this end, we used (\ref{v_value}) to eliminate the parameter $v$ and expand the function given in (\ref{eq82}) in powers of the weak mixing angle. This yields (with $s_W^2\equiv\sin^2\theta_W$)
\begin{equation}
   \Gamma(Z\to M^+W^-) 
   = \frac{\pi\alpha^2(m_Z)\,f_M^2}{48 m_Z}\,|V_{ij}|^2\,\frac{s_W^2}{c_W^2}
    \left( \frac{3}{2} + \frac{3}{2}\,s_W^2 + \frac{227}{180}\,s_W^4 
    + 0.003\,a_1^M + \dots \right) ,
\end{equation}
where we do not show terms of ${\cal O}(s_W^6)$ and contributions from higher Gegenbauer moments. The leading contributions in this expression are in agreement with those found in \cite{Manohar:1990hu}, where an anlogous formula with the parenthesis replaced by $\frac{6}{(1+c_W^2)^2}=\frac{3}{2}+\frac{3}{2}\,s_W^2+\frac{9}{8}\,s_W^4+\dots$ is given.

\section{Experimental considerations}
\label{sec:experiments}

Having obtained detailed and accurate theoretical predictions for a large class of very rare, exclusive radiative decays of $Z$ and $W$ bosons, we now address the question of how to search for such decays experimentally. While we do not perform an exhaustive feasibility study here, we discuss several ideas related to possible experimental analysis strategies. The goal is to get a feeling for how difficult it will be to observe some of the decay modes discussed in this work at present and future particle accelerators, and what accuracy one will be able to reach. We first consider the LHC, where about $10^{11}$ $Z$ bosons and $5\cdot 10^{11}$ $W$ bosons will have been produced in both ATLAS and CMS by the time the high-luminosity run with an anticipated integrated luminosity of 3000\,fb$^{-1}$ has been completed. We also consider a future lepton collider such as TLEP \cite{Blondel:2013rn}, where one can hope to produce about $10^{12}$ $Z$ bosons per year with a dedicated run on the $Z$ pole, while $10^7$ $W$-boson pairs per year would be produced in a run at the $WW$ threshold. A run just above the $t\bar t$ threshold would also produce very large $WW$ samples.

At the LHC one needs to worry about triggering and reconstruction. The trigger for the decays $Z\to M\gamma$ can be based on photons and muons. The energy of the final-state photon is comparable to the energy of the photons produced in the Higgs-boson decay $h\to\gamma\gamma$, where it has been demonstrated that such events can be triggered on. Muons, on the other hand, are produced only in some cases, in particular in the decay modes containing a vector meson. Reconstructing these event appears to be challenging but not impossible. Probably the most promising modes are $Z\to J/\psi\,\gamma$ and $Z\to\Upsilon(nS)\,\gamma$ followed by a fully leptonic decay of the heavy quarkonium state. The corresponding rates, however, are very small. If only the muon channel can be used, then the combined $Z\to V\gamma\to\mu^+\mu^-\gamma$ branching fractions are of order $5\cdot 10^{-9}$ for $J/\psi$ and $1.5\cdot 10^{-9}$ for $\Upsilon(1S)$. Thus, we can expect to trigger on several hundred $Z\to J/\psi\,\gamma\to\mu^+\mu^-\gamma$ events and up to one hundred $Z\to \Upsilon(1S)\,\gamma\to\mu^+\mu^-\gamma$ events at the LHC. While there is no significant physics background we can think of, the combinatorial background may be substantial. Given these challenges, it is encouraging that ATLAS has recently reported first upper bounds (at 95\% CL) on the branching fractions $\mbox{Br}(Z\to J/\psi\,\gamma)<2.6\cdot 10^{-6}$, $\mbox{Br}(Z\to\Upsilon(1S)\,\gamma)<3.4\cdot 10^{-6}$, $\mbox{Br}(Z\to\Upsilon(2S)\,\gamma)<6.5\cdot 10^{-6}$, and $\mbox{Br}(Z\to\Upsilon(3S)\,\gamma)<5.4\cdot 10^{-6}$ \cite{Aad:2015sda}. Further dedicated experimental studies of these decays would be worthwhile. The yield of $Z\to\Upsilon\gamma$ events can be enhanced by about a factor 2 if one combines the $\Upsilon(nS)$ channels with $n=1,2,3$, which have similar leptonic branching fractions (ranging between 1.9\% and 2.5\%). The case $Z\to\Upsilon(4S)\,\gamma$ may be particularly interesting, since the $\Upsilon(4S)$ resonance can decay to a pair of $B$ mesons, which gives rise to displaced vertices. It might be possible to achieve a larger effective rate for this decay mode by using highly efficient $b$-tagging methods. Observing $Z\to M\gamma$ decays into other final states seems to be difficult at the LHC. Ideas for reconstructing highly energetic $\phi$, $\rho$ and $\omega$ mesons have been presented in \cite{Kagan:2014ila} in the context of a study of exclusive $h\to V\gamma$ decays. For light mesons decaying into two photons, such as $\pi^0$ and $\eta$, it might be possible to tell that there are more than one photon in the final state provided one of the photons is converted into an $e^+ e^-$ pair. 

The situation with $W\to M\gamma$ decays at the LHC seems less promising. In this case the final state contains a charged hadron, and it is not clear to us how one could reconstruct it in the high-multiplicity environment of a hadron collider. Perhaps the most promising case is the decay $W\to D_s\gamma$, of which over 10,000 events should be produced. The problem is how to tag the $D_s$ mesons. A very interesting, dedicated study of several exclusive radiative $W\to M\gamma$ decays has been performed in \cite{Mangano:2014xta}, to which we refer the reader for more details. 

It looks much more promising to search for rare exclusive $Z\to M\gamma$ decays at a future lepton collider such as TLEP, in particular if one envisions a dedicated high-luminosity run on the $Z$ pole. The advantage of such a ``$Z$ factory'' is that it would produce yields of up to $10^{12}$ $Z$ bosons per year in a very clean environment. The detectors of such a future experiment are expected to have excellent particle-identification systems. This would make it possible to perform precision studies of many of the decay modes we have discussed. In particular, all $Z\to M\gamma$ decays except for $Z\to\pi^0\gamma$ should be accessible, and in several cases it should be possible to measure the branching ratios at the percent level. There is even hope for observing some of the weak radiative decays $Z\to MW$, whose branching fractions can be of order several times $10^{-10}$. With dedicated runs above the $WW$ or $t\bar t$ thresholds one would produce samples of $W$ bosons large enough to search for the Cabibbo-allowed $W\to M\gamma$ decay modes. It would be most rewarding to perform detailed feasibility studies for measurements of rare  exclusive $Z$ and $W$ decays at such a facility.

\section{Summary and conclusions}
\label{sec:concl}

Based on the formalism of QCD factorization, we have performed a comprehensive and systematic analysis of the very rare, exclusive radiative decays of $Z$ and $W$ bosons into final states containing a single pseudoscalar or vector meson. The basis of our study is a factorization theorem derived in soft-collinear effective theory, which expresses the decay amplitudes as convolutions of calculable hard-scattering kernels with light-cone distribution amplitudes (LCDAs), in a systematic expansion in powers of $(\Lambda_{\rm QCD}/m_{Z,W})^2$ and $(m_M/m_{Z,W})^2$. For the first time, we have included the complete set of one-loop QCD radiative corrections. Large logarithms involving the ratio of the electroweak scale to the typical hadronic scale $\mu_0$, at which model predictions for the LCDAs are obtained, are resummed to all orders of perturbation theory. We have also estimated the leading power-suppressed effects, finding that their impact on the branching ratios is typically of order $10^{-4}$ compared with the leading terms. Larger power corrections up of to 1\% can only arise for mesons containing heavy $b$ quarks. The exclusive decays $Z\to M\gamma$ and $W\to M\gamma$ therefore offer an ideal laboratory for testing the QCD factorization approach in a controlled and theoretically clean way. Our main phenomenological results for the relevant branching ratios are collected in Table~\ref{tab:BRsSummary}. We have not considered in this work the interesting decays $Z\to\eta^{(\prime)}\gamma$. Their analysis is complicated by the fact that the $\eta$ and $\eta'$ mesons contain a flavor-singlet Fock component, and to predict the decay rates at ${\cal O}(\alpha_s)$ one needs to take into account the two-gluon LCDA of these mesons. This will be discussed in a separate publication \cite{inprep}.

\begin{table}
\begin{center}
\begin{tabular}{|c|c||c|c|}
\hline 
Decay mode & Branching ratio & Decay mode & Branching ratio \\ 
\hline 
$Z^0\to\pi^0\gamma$ & $(9.80\pm 1.03)\cdot 10^{-12}$ &
 $W^\pm\to\pi^\pm\gamma$ & $(4.00\pm 0.83)\cdot 10^{-9}$ \\ 
$Z^0\to\rho^0\gamma$ & $(4.19\pm 0.47)\cdot 10^{-9}$ &
 $W^\pm\to\rho^\pm\gamma$ & $(8.74\pm 1.91)\cdot 10^{-9}$ \\ 
$Z^0\to\omega\gamma$ & $(2.82\pm 0.41)\cdot 10^{-8}$ &
 $W^\pm\to K^\pm\gamma$ & $(3.25\pm 0.69)\cdot 10^{-10}$ \\ 
$Z^0\to\phi\gamma$ & $(1.04\pm 0.12)\cdot 10^{-8}$ &
 $W^\pm\to K^{*\pm}\gamma$ & $(4.78\pm 1.15)\cdot 10^{-10}$ \\ 
$Z^0\to J/\psi\,\gamma$ & $(8.02\pm 0.45)\cdot 10^{-8}$ & 
 $W^\pm\to D_s\gamma$ & $(3.66\,_{-\,0.85}^{+\,1.49})\cdot 10^{-8}$ \\ 
$Z^0\to\Upsilon(1S)\,\gamma$ & $(5.39\pm 0.16)\cdot 10^{-8}$ & 
 $W^\pm\to D^\pm\gamma$ & $(1.38\,_{-\,0.33}^{+\,0.51})\cdot 10^{-9}$ \\ 
$Z^0\to\Upsilon(4S)\,\gamma$ & $(1.22\pm 0.13)\cdot 10^{-8}$ &
 $W^\pm\to B^\pm\gamma$ & $(1.55\,_{-\,0.60}^{+\,0.79})\cdot 10^{-12}$ \\
\hline 
\end{tabular}
\parbox{15.5cm}
{\caption{\label{tab:BRsSummary} 
Summary table of our predictions for the branching fractions of exclusive radiative decays of $Z$ and $W$ bosons. Different sources of theoretical errors have been added in quadrature.}}
\end{center}
\end{table} 

Our results form the basis of a rich, novel program of electroweak precision studies of the Standard Model, which offers powerful tests of the QCD factorization approach in a situation where it should deliver precise predictions. With the statistics obtainable in the high-luminosity run at the LHC and, in a much cleaner environment, at future lepton colliders, it will be possible to observe several of the very rare decays discussed here and measure their branching fractions with some accuracy. Precise rate measurements, which would be possible at a dedicated $Z$-boson factory, would offer the unique possibility to extract highly non-trivial information about the LCDAs of various mesons in a completely model-independent way. More specifically, for each meson $M$ one will be able to extract the sums over the even and odd Gegenbauer moments, $\sum_n a_{2n}^M(\mu)$ and $\sum_n a_{2n+1}^M(\mu)$, at the electroweak scale $\mu\sim m_Z$, up to small and calculable radiative corrections. This is a consequence of the structure of the basic convolution integrals in (\ref{Ipmdef}). We cannot imagine a theoretically cleaner way to get access to this kind of information. We have also performed an exploratory study of the weak radiative decays $Z\to MW$, which allow for tests of the QCD factorization approach at lower scales $\mu\sim 10$\,GeV, which are only a few times higher than those relevant to exclusive hadronic $B$-meson decays. Our predictions for the corresponding branching fractions obtained at tree level have been given in Table~\ref{tab:BRsZWM}.

Several generalizations and extensions of our work are possible and worth exploring. Our formalism can be applied in a straightforward way to obtain high-precision predictions for exclusive radiative (and weak radiative) decays of the Higgs boson, extending previous tree-level analyses presented in \cite{Isidori:2013cla,Bodwin:2013gca,Kagan:2014ila,Bodwin:2014bpa,Gao:2014xlv,Bhattacharya:2014rra}. One goal of such studies is to search for enhanced Yukawa couplings and flavor-changing interactions of the Higgs boson. In this context, new-physics studies analogous to those presented in Section~\ref{sec:numerics} are particularly interesting. Without further conceptual developments, our formalism can also be extended to calculate the rates for purely hadronic decays, such as $Z,W,h\to M_1 M_2$ or even decays with more than two particles in the final state. These extensions are left for future work.

The physics case for studying some of the very rare, exclusive decays of heavy electroweak bosons is compelling to us. There is some beautiful physics to be explored here, both from the theoretical and the experimental points of view. We hope that our detailed exploratory survey will raise sufficient interest that some dedicated feasibility studies for discovering such decays at the high-luminosity LHC and future lepton colliders will be performed.

\subsubsection*{Acknowledgments}

We thank Alexander Kagan for useful discussions, which motivated us to start this research. We are grateful to Volodya Braun, Jens Erler and Gilad Perez for valuable comments. The work of Y.G.\ is supported is part by the U.S.\ National Science Foundation through grant PHY-0757868 and by the United States-Israel Binational Science Foundation (BSF) under grant no.~2010221. The work of M.K.\ and M.N.\ is supported by the Advanced Grant EFT4LHC of the European Research Council (ERC), the Cluster of Excellence {\em Precision Physics, Fundamental Interactions and Structure of Matter\/} (PRISMA -- EXC 1098), grant 05H12UME of the German Federal Ministry for Education and Research (BMBF), and the DFG Graduate School {\em Symmetry Breaking in Fundamental Interactions\/} (GRK 1581).

\begin{appendix}

\section{Light-cone projectors for vector mesons}
\label{app:LCDAs}
\renewcommand{\theequation}{A.\arabic{equation}}
\setcounter{equation}{0}

The LCDAs of vector mesons at leading and subleading twist have been studied in great detail in \cite{Ali:1993vd,Ball:1996tb,Ball:1998sk}. The corresponding momentum-space projectors were derived in \cite{Beneke:2000wa}. In analogy with (\ref{LCDAP}), the light-cone projector at leading and subleading power for a longitudinally polarized vector meson reads
\begin{equation}\label{LCDAVlong}
\begin{aligned}
   M_{V_\parallel}(k,x,\mu) &= - \frac{if_V}{4}\,\rlap{\hspace{0.1mm}/}{k}\,\phi_V(x,\mu) 
    - \frac{if_V^\perp(\mu)\,m_V}{4}\,\Bigg\{ \frac{h_\parallel^{\prime\,(s)}(x,\mu)}{2}
    - i\sigma_{\mu\nu}\,\frac{k^\mu\,\bar n^\nu}{k\cdot\bar n}\,h_\parallel^{(t)}(x,\mu) \\[-1mm]
   &\hspace{5mm}\mbox{}- i\sigma_{\mu\nu} k^\mu\,\int_0^x\!dy 
    \left[ \phi_V^\perp(y,\mu) - h_\parallel^{(t)}(y,\mu) \right] 
    \frac{\partial}{\partial k_{\perp\nu}} + \mbox{3-particle LCDAs} \Bigg\} \,.
\end{aligned}
\end{equation}
In the approximation where three-particle LCDAs are neglected, the QCD equations of motion imply the relations \cite{Ball:1998sk,Beneke:2000wa}
\begin{equation}
\begin{aligned}
   h_\parallel^{(t)}(x,\mu) \big|_{\rm WWA} &= (2x-1)\,\Phi_v(x,\mu) \,, \qquad
    h_\parallel^{\prime\,(s)}(x,\mu) \big|_{\rm WWA} = -2\Phi_v(x,\mu) \,, \\
   &\hspace{-12mm} \int_0^x\!dy \left[ \phi_V^\perp(y,\mu) - h_\parallel^{(t)}(y,\mu) 
    \right]_{\rm WWA} = x(1-x)\,\Phi_v(x,\mu) \,,
\end{aligned}
\end{equation}
where
\begin{equation}
   \Phi_v(x,\mu) = \int_0^x\!dy\,\frac{\phi_V^\perp(y,\mu)}{1-y} 
    - \int_x^1\!dy\,\frac{\phi_V^\perp(y,\mu)}{y} \,.
\end{equation}
In this approximation, the twist-3 two-particle amplitudes can be expressed in terms of the twist-2 LCDA $\phi_V^\perp$.

The light-cone projector for a transversely polarized vector meson is yet more complicated. Up to twist-3 order, one obtains 
\begin{equation}\label{LCDAVperp}
\begin{aligned}
   M_{V_\perp}(k,x,\mu) &= \frac{if_V^\perp(\mu)}{4}\,\rlap{\hspace{0.1mm}/}{k}\,
    \rlap/\varepsilon_V^{\perp *}\,\phi_V^\perp(x,\mu) 
    - \frac{if_V m_V}{4} \Bigg\{ \rlap/\varepsilon_V^{\perp *}\,g_\perp^{(v)}(x,\mu) \\
   &\quad\mbox{}- \frac{i}{4}\,\epsilon_{\mu\nu\alpha\beta}\,\gamma^\mu\gamma_5\,
    \varepsilon_V^{\perp *\nu} k^\alpha 
    \left( \frac{\bar n^\beta}{k\cdot\bar n}\,g_\perp^{\prime\,(a)}(x,\mu)
    - g_\perp^{(a)}(x,\mu)\,\frac{\partial}{\partial k_{\perp\beta}} \right) \\
   &\quad\mbox{}- \rlap{\hspace{0.1mm}/}{k}\,\varepsilon_{V\mu}^{\perp *}
    \int_0^x\!dy \left[ \phi_V(y,\mu) - g_\perp^{(v)}(y,\mu) \right]
    \frac{\partial}{\partial k_{\perp\mu}} + \mbox{3-particle LCDAs} \Bigg\} \,.
\end{aligned}
\end{equation}
Note that, compared with \cite{Beneke:2000wa}, the terms multiplying the Levi-Civita tensor in the second line have the opposite sign, because we are using a different sign convention for this object. In the approximation where three-particle LCDAs are neglected, the equations of motion yield the relations \cite{Ball:1996tb,Ball:1998sk,Beneke:2000wa}
\begin{equation}\label{WWArela}
\begin{aligned}
   g_\perp^{(v)}(x,\mu) \big|_{\rm WWA} 
   &= \frac12 \left[ \int_0^x\!dy\,\frac{\phi_V(y,\mu)}{1-y} 
    + \int_x^1\!dy\,\frac{\phi_V(y,\mu)}{y} \right] , \\
   g_\perp^{(a)}(x,\mu) \big|_{\rm WWA} 
   &= 2 \left[ (1-x) \int_0^x\!dy\,\frac{\phi_V(y,\mu)}{1-y} 
    + x \int_x^1\!dy\,\frac{\phi_V(y,\mu)}{y} \right] , \\
   g_\perp^{\prime\,(a)}(x,\mu) \big|_{\rm WWA} 
   &= -2 \left[ \int_0^x\!dy\,\frac{\phi_V(y,\mu)}{1-y} 
    - \int_x^1\!dy\,\frac{\phi_V(y,\mu)}{y} \right] , \\
   \int_0^x\!dy \left[ \phi_V(y,\mu) - g_\perp^{(v)}(y,\mu) \right]
   &= \frac12 \left[ (1-x) \int_0^x\!dy\,\frac{\phi_V(y,\mu)}{1-y} 
    - x \int_x^1\!dy\,\frac{\phi_V(y,\mu)}{y} \right] .
\end{aligned}
\end{equation}
In this approximation, the twist-3 two-particle amplitudes can be expressed in terms of the twist-2 LCDA $\phi_V$.

As an application of these results, we present the general expressions for the form factors $F_i^\perp$ entering the $Z\to V_\perp\gamma$ and $W^+\to V_\perp^+\gamma$ decay amplitudes in (\ref{amplperp}) and (\ref{amplperp2}). In the first case, we obtain (with $\bar x\equiv 1-x$, and ignoring quark-mass effects for simplicity)
\begin{equation}
\begin{aligned}
   F_1^\perp &= \frac{{\cal Q}_V}{6}\,\Bigg\{ 
    \int_0^1\!dx \left( \frac{1+x}{x} + \frac{1+\bar x}{\bar x} \right) g_\perp^{(v)}(x) 
    - \int_0^1\!dx \left( \frac{1-x}{x} - \frac{1-\bar x}{\bar x} \right)
    \frac{g_\perp^{\prime\,(a)}(x)}{4} \\
   &\hspace{15mm}\mbox{}+ \int_0^1\!dx \left( \frac{1}{x} + \frac{1}{\bar x} \right) 
    \frac{g_\perp^{(a)}(x)}{4} 
    + \int_0^1\!dx \left( \frac{1}{x} - \frac{1}{\bar x} \right) 
    \int_0^x\!dy \left[ \phi_V(y) - g_\perp^{(v)}(y) \right] \Bigg\} \,, \\
   F_2^\perp &= \frac{{\cal Q}_V'}{6}\,\Bigg\{ 
    \int_0^1\!dx \left( \frac{1-x}{x} - \frac{1-\bar x}{\bar x} \right) g_\perp^{(v)}(x) 
    - \int_0^1\!dx \left( \frac{1+x}{x} + \frac{1+\bar x}{\bar x} \right)
    \frac{g_\perp^{\prime\,(a)}(x)}{4} \\
   &\hspace{15mm}\mbox{}+ \int_0^1\!dx \left( \frac{1}{x} - \frac{1}{\bar x} \right) 
    \frac{g_\perp^{(a)}(x)}{4} 
    + \int_0^1\!dx \left( \frac{1}{x} + \frac{1}{\bar x} \right) 
    \int_0^x\!dy \left[ \phi_V(y) - g_\perp^{(v)}(y) \right] \Bigg\} \,.
\end{aligned}
\end{equation}
We omit the scale dependence of the various quantities for simplicity. In the case of (\ref{amplperp2}) we must omit the prefactors ${\cal Q}_V^{(\prime)}/6$ and instead assign charge factors $Q_u$ and $Q_d$ to the two terms under each integral over $x$. These results can be simplified significantly by using the relations (\ref{WWArela}). The final expressions have been given in (\ref{HAVperpres}) and (\ref{Hperpres2}).

\section{Determinations of meson decay constants}
\label{app:decay_constants}
\renewcommand{\theequation}{B.\arabic{equation}}
\setcounter{equation}{0}

We follow \cite{Neubert:1997uc} and determine the relevant meson decay constants from experimental data. The decay constants of charged pseudoscalar mesons can be determined from their semileptonic decays $P^-\to l^-\bar\nu_l(\gamma)$. This analysis is performed by the Particle Data Group \cite{Agashe:2014kda}, and it leads to the values for $f_\pi$, $f_K$, $f_D$ and $f_{D_s}$ shown in Tables~\ref{tab:hadronic_inputs} and \ref{tab:hadronic_inputs2}. 

The decay constants of light charged mesons can also be obtained from the one-prong hadronic decays of the $\tau$ lepton. The corresponding decay rates are given by
\begin{equation}
   \Gamma(\tau^-\to M^-\nu_\tau) = S_{\rm EW}\,\frac{G_F^2 m_\tau^3}{16\pi}\,|V_{ij}|^2 f_M^2
    \left( 1 - \frac{m_M^2}{m_\tau^2} \right)^2 \!
    \left( 1 + b_M\,\frac{m_M^2}{m_\tau^2} \right) ,
\end{equation}
where $b_P=0$ for pseudoscalar mesons and $b_V=2$ for vector mesons. $V_{ij}$ are the relevant CKM matrix elements. The factor $S_{\rm EW}=1.0154$ includes the leading-logarithmic \cite{Sirlin:1977sv,Marciano:1988vm} and non-logarithmic electroweak corrections \cite{Braaten:1990ef}. From the measured branching fractions $\mbox{Br}(\pi^-)=(10.83\pm 0.06)\%$, $\mbox{Br}(K^-)=(0.70\pm 0.01)\%$, $\mbox{Br}(\rho^-)=(25.22\pm 0.33)\%$ and $\mbox{Br}(K^{*-})=(1.20\pm 0.07)\%$, along with the $\tau$-lepton lifetime $\tau_\tau=(290.3\pm 0.5)\cdot 10^{-15}$\,s \cite{Agashe:2014kda}, we extract $f_\pi=(130.3\pm 0.4)$\,MeV, $f_K=(154.3\pm 1.1)$\,MeV, $f_\rho=(207.8\pm 1.4)$\,MeV, and $f_{K^*}=(203.2\pm 5.9)$\,MeV. The values of $f_\pi$ and $f_K$ are in excellent agreement with the (more precise) values extracted from semileptonic decays, supporting the reliability of this method. 

The decay constants of neutral vector mesons can be extracted from their electromagnetic decay width using
\begin{equation}
   \Gamma(V^0\to e^+ e^-) = \Gamma(V^0\to\mu^+\mu^-)
   = \frac{4\pi f_V^2}{3 m_V}\,\alpha^2(m_V)\,c_V \,,
\end{equation}
where the coefficients $c_V=\big(\sum_q c_q^V Q_q\big)^2$ are related to the electric charges of the quarks that make up the vector meson \cite{Neubert:1997uc}. Explicitly, one has $c_\rho=1/2$, $c_\omega=1/18$, $c_\phi=1/9$, $c_{J/\psi}=4/9$, and $c_\Upsilon=1/9$. The dominant QED corrections are accounted for by using the electromagnetic coupling evaluated at $\mu=m_V$, which we compute using $\alpha(m_Z)^{-1}=127.94$ and the approach described in \cite{Erler:1998sy}. Averaging over the $e^+ e^-$ and $\mu^+\mu^-$ modes \cite{Agashe:2014kda}, we obtain the measured branching fractions $\mbox{Br}(\rho^0)=(4.715\pm 0.049)\cdot 10^{-5}$, $\mbox{Br}(\omega)=(7.284\pm 0.140)\cdot 10^{-5}$, $\mbox{Br}(\phi)=(2.952\pm 0.030)\cdot 10^{-4}$, and when combined with the total widths $\Gamma(\rho^0)=(147.8\pm 0.9)$\,MeV, $\Gamma(\omega)=(8.49\pm 0.08)$\,MeV, $\Gamma(\phi)=(4.266\pm 0.031)$\,MeV this yields the decay constants $f_\rho=(216.3\pm 1.3)$\,MeV, $f_\omega=(194.2\pm 2.1)$\,MeV and $f_\phi=(223.0\pm 1.4)$\,MeV. For the heavy quarkonium states it is advantageous to use the measured electromagnetic width directly. They are $\Gamma_{ee}(J/\psi)=(5.55\pm 0.14)$\,keV, $\Gamma_{ee}(\Upsilon(1S))=(1.340\pm 0.018)$\,keV, $\Gamma_{ee}(\Upsilon(2S))=(0.612\pm 0.011)$\,keV, $\Gamma_{ee}(\Upsilon(3S))=(0.443\pm 0.008)$\,keV and $\Gamma_{ee}(\Upsilon(4S))=(0.272\pm 0.029)$\,keV \cite{Agashe:2014kda}. This gives the decay constants $f_{J/\psi}=(403.3\pm 5.1)$\,MeV, $f_{\Upsilon(1S)}=(684.4\pm 4.6)$\,MeV, $f_{\Upsilon(2S)}=(475.8\pm 4.3)$\,MeV, $f_{\Upsilon(3S)}=(411.3\pm 3.7)$\,MeV and $f_{\Upsilon(4S)}=(325.7\pm 17.4)$\,MeV. For our analysis the value $(\sum_{n=1}^3 f_{\Upsilon(nS)}^2)^{1/2}=(930\pm 4)$\,MeV is also needed. As discussed in \cite{Ball:2006eu},\footnote{We are grateful to R.~Zwicky for pointing out a numerical mistake in the published version of this paper.} 
the combined effects of $\rho$\,--\,$\omega$ and $\omega$\,--\,$\phi$ mixing lower $f_\omega$ by about 9.5\,MeV and raise $f_\phi$ by about 7.6\,MeV, while they have a negligible impact on $f_\rho$. Note also that there are some uncertainties related to the sizable width of the $\rho$ resonance, which we have ignored here. They might explain why the value of $f_{\rho^0}$ extracted here is larger than the value of $f_{\rho^-}$ extracted above from $\tau$ decay. Combining the various extractions and using conservative error estimates, we obtain the values shown in Table~\ref{tab:hadronic_inputs}.

\section{Gegenbauer expansion of the convolution integrals}
\label{app:moments}
\renewcommand{\theequation}{C.\arabic{equation}}
\setcounter{equation}{0}

In order to derive closed analytic expressions for the basic convolution integrals $I_\pm^M$ and $\bar I_\pm ^M$ defined in (\ref{Ipmdef}), we employ the definition of the Gegenbauer polynomials in terms of the generating function
\begin{equation}
   \frac{1}{\left(1-2xt+t^2\right)^\alpha} = \sum_{n=0}^\infty\,C_n^{(\alpha)}(x)\,t^n \,.
\end{equation} 
From (\ref{Gegenbauer}) and the second relation in (\ref{Ipmdef}), it follows that the coefficients $C_n^{(\pm)}(m_V,\mu)$ defined in (\ref{Cndef}) are the coefficients of $t^n$ of the integrals
\begin{equation}
   \int_0^1\!dx\,\frac{2x(1-x)}{\left[(1+t)^2-4xt\right]^{3/2}}\,H_\pm(1-x,m_V,\mu) 
   = \frac{1}{1-t} + \frac{C_F\alpha_s(\mu)}{4\pi}\,h(t) + {\cal O}(\alpha_s^2) \,,
\end{equation}
where
\begin{equation}
\begin{aligned}
   h(t) &= \left[ \frac{2}{t^2} \left( \frac{1+t^2}{1-t} \ln(1-t) + t \right) 
    + \frac{3}{1-t} \right] \left( \ln\frac{m_V^2}{\mu^2} - i\pi \right) - \frac{9}{1-t} \\
   &\quad\mbox{}+ \frac{2(1+t^2)}{t^2(1-t)}
    \left[ \ln^2(1-t) + \mbox{Li}_2(t) \right] 
    - \frac{2}{t} \mp \left[ \frac{t + (1-t)\ln(1-t)}{t^2} \right] .
\end{aligned}
\end{equation}
The coefficients $c_n^{(\pm)}(m_V/\mu)$ in (\ref{Cndef}) are the coefficients of $t^n$ in the series expansion of $h(t)$ around $t=0$. It is then a straightforward exercise to derive expression (\ref{gorgeous}).

\section{Connection with the approach by Manohar}
\label{app:manohar}
\renewcommand{\theequation}{D.\arabic{equation}}
\setcounter{equation}{0}

The approach put forward by Manohar in \cite{Manohar:1990hu} was originally developed as a tool to study the exclusive decay $Z\to W\pi$ by means of a local operator-product expansion. It corresponds to the following series expansion of the propagator in the first diagram in Figure~\ref{fig:diags}:
\begin{equation}
   \frac{1}{x m_Z^2+(1-x) m_W^2}
   = \frac{2}{m_Z^2+m_W^2}\,\frac{1}{\left[ 1-\left(\frac12-x\right) \omega_0 \right]} 
   = \frac{2}{m_Z^2+m_W^2}\,\bigg[ 1 + \sum_{n=0}^\infty\Big( \frac{\omega_0}{2} \Big)^n 
    (1-2x)^n \bigg] \,, 
\end{equation}
where
\begin{equation}
   \frac{\omega_0}{2} = \frac{m_Z^2-m_W^2}{m_Z^2+m_W^2} \approx 0.125 \,.
\end{equation}
Using this expansion, the convolution integrals over the pion LCDA in (\ref{ZWMffs}) can be expressed in terms of local operator matrix elements, and the leading terms are determined by the normalization of the LCDA. For the phenomenological estimates in \cite{Manohar:1990hu} and \cite{Mangano:2014xta} only the leading term was kept. 

When the same approach is used for radiative decays, the $W$-boson mass in the above relations must be set to zero, in which case one obtains
\begin{equation}
   \frac{1}{x m_Z^2}
   = \frac{2}{m_Z^2}\,\bigg[ 1 + \sum_{n=0}^\infty (1-2x)^n \bigg] \,. 
\end{equation}
Keeping the leading term is now unjustified, but if it is done this corresponds to replacing $1/x\to 2$ under the convolution integrals in (\ref{Ipmdef}). In the asymptotic limit, where all Gegenbauer moments are set to zero, these integrals are therefore too small by a factor 2/3. This explains the discrepancies in the predictions for the $Z\to M\pi$ and $W\to M\pi$ branching fractions mentioned in Section~\ref{sec:numerics}.

\end{appendix}

\end{document}